\documentclass[review,onefignum,onetabnum]{article}

\usepackage[a4paper,
bindingoffset=0.2in,
left=1in,
right=1in,
top=1in,
bottom=1in,
footskip=.25in]{geometry}
\usepackage{float}
\usepackage{flafter}

\usepackage[dvips]{epsfig}
\usepackage{mathtools}
\usepackage{natbib}
\usepackage{xcolor}
\usepackage{dsfont}
\usepackage{tikz}
\usepackage{pict2e}

\usepackage{amsmath}
\usepackage[utf8]{inputenc}
\usepackage[english]{babel}
\usepackage{graphicx}
\usepackage{comment}
\usepackage{epic}
\usepackage{lscape}
\usepackage{rotating}
\usepackage{epstopdf}
\usepackage{wrapfig}
\usepackage{amssymb}
\usepackage{bm}

\usepackage{booktabs}

\usepackage{array}

\newcommand\MCC[0]
{%
	\ensuremath
	{
		\text{MCC3}
	}
}
\newcommand\lrr[1]
{%
	\ensuremath
	{
		\left( #1 \right)
	}
}

\usepackage{overpic}

\usepackage{soul}

\title{Large mode-2 internal solitary waves in three-layer flows }
\author{Alex Doak\thanks{Corresponding author. Department of Mathematical Sciences, University of Bath, UK BA2 2AY. add49@bath.ac.uk}
	\and Ricardo Barros\thanks{ Department of Mathematical Sciences, Loughborough university, Loughborough LE11 3TU,
		UK }
	\and Paul A Milewski \thanks{ Department of Mathematical Sciences, University of Bath, UK BA2 2AY}}

\date{30/04/2022}

\begin{document}
	\maketitle
\begin{abstract}
In this paper, we investigate mode-2 solitary waves in a three-layer stratified flow model. Localised travelling wave solutions to both the fully nonlinear problem (Euler equations), and the three-layer Miyata-Choi-Camassa equations are found numerically and compared. Mode-2 solitary waves with speeds slower than the linear mode-1 long-wave speed are typically generalised solitary waves with infinite tails consisting  of a resonant mode-one periodic wave train. Herein we evidence the existence of mode-2 embedded solitary waves, that is, we show that for specific values of the parameters, the amplitude of the oscillations in the tail are zero. For sufficiently thick middle layers, we also find branches of mode-2 solitary waves with speeds that extend beyond the mode-1 linear waves and are no longer embedded.
In addition, we show how large amplitude embedded solitary waves  are intimately linked to the conjugate states of the problem.
\end{abstract}

\section{Introduction}

Waves propagating in density stratified fluids are responsible for mass, momentum, temperature, and biomass transfer in the world's oceans. The ocean is a continuously stratified fluid, for which linear theory predicts infinitely many modes of propagation.  
Most previous literature, both observational and theoretical, concerns waves of the  first baroclinic mode (mode-1), which have the simplest vertical structure.  Recent observations
suggest that mode-2 waves are more common than previously thought \citep{duda2004internal,yang2009observations,shroyer2010mode, ramp2012observations,khimchenko2016mode}. Although observations by \cite{shroyer2010mode} found mode-2 waves off the coast of New Jersey were 10 to 100 times less energetic than their mode-1 counterparts,  they are capable of  travelling large distances with  so-called `trapped-cores', inducing mass transport \citep{brandt2014laboratory,deepwell2016mass}. Furthermore, \cite{shroyer2010mode} shows that mode-2 waves induce similar measures of localised turbulent kinetic energy dissipation as mode-1 waves; as such they have considerable influence on vertical mixing, vertical transport of heat and nutrients and ultimately are important in climate models. Mode-2 waves have also been the subject of experimental studies  (for example, \citeauthor{kao1980wake} \citeyear{kao1980wake};
\citeauthor{stamp1995deep} \citeyear{stamp1995deep}; \citeauthor{honji1995experimental} \citeyear{honji1995experimental}; \citeauthor{carr2015experiments} \citeyear{carr2015experiments}).

Herein, we explore strongly nonlinear effects on mode-2 waves in a three-layer rigid-lid configuration. The rigid-lid approximation is commonly used when studying weakly stratified fluids such as the ocean, and replaces the free-surface with a solid wall \citep{evans1996integral}. Layered models with layers of constant density are a common approximation in settings where the physical stratification profile is composed of one or more regions of almost constant density separated by regions of rapid density change (pycnoclines). For example \cite{grue1999properties} compared steady travelling mode-1 ISWs of the two-layer Euler equations against experiments conducted using salted water with one pycnocline. It was found that the layered model yielded good approximations for wave profiles and the velocity field in the water column across the whole amplitude range.

Various model equations are often used to describe internal waves, most notably the Korteweg-de Vries (KdV) equation which is meant to capture well weakly nonlinear shallow water phenomena \citep{benney1966long,grimshaw2003internal}.  \cite{miyata1988long} and \cite{choi1999fully} independently derived a model system for a two-layered fluid (henceforth denoted MCC2) which makes no assumption on the nonlinearity but assumes shallow water. 
\cite{camassa2006realm} compared steady travelling waves of the MCC2 system against those of KdV, of the two-layered Euler equations, and the experiments of \cite{grue1999properties} and \cite{sveen2002breaking}. They found the MCC2 system to be in superb agreement with the Euler equations and experiments in both wave profiles and velocity fields, except in regions where the shallow water approximation is invalidated. The model was later generalised for an arbitrary number of layers \citep{choi2000modeling}. We shall refer to the two- and three-layer Miyata-Choi-Camassa equations as MCC2 and MCC3 respectively.

Linear and weakly nonlinear theories predict that both interface displacements have the same polarity for mode-1, while mode-2 waves have interfaces with opposite polarities. If the lower interface is of depression (and hence the upper interface is of elevation), then the wave is called  convex. The opposite case, that is of a lower interface of elevation, is known as concave. Whether mode-2 solitary waves are convex or concave can be predicted by the weakly nonlinear KdV equation, where it is found that thicker middle layers result in concave waves. 

Typically, mode-2 nonlinear waves have speeds coinciding with that of shorter mode-1 waves, thus mode-2 solitary wave speeds lie within the linear spectrum of the system.
Hence, due to resonance with a mode-1 short wave travelling with the same speed, they are expected to have  non-decaying oscillatory mode-1 tails. Such solutions are called generalised solitary waves (GSW) and have infinite energy. However, it could happen that solitary wave solutions within the linear spectrum have a flat far-field (i.e. no resonant tail) and are ``true" solitary waves. In other contexts, these were coined embedded solitary waves (ESW) by \cite{yang1999embedded} as their speed is embedded within the linear spectrum. One may find these solutions as particular cases of GSW, for which the amplitude of the oscillatory tail goes to zero, such as the ESWs discovered for the modified fifth-order KdV equation by \cite{champneys2002true}.

While mode-2 ESWs have been found before in the context of the stratified Euler equations, all examples so far are restricted to the highly specific construction of reflecting a mode-1 ISW across an imaginary wall at the mid-channel  \citep{davis1967solitary,tung1982large,king2011steady}. This construction prevents a mode-1 tail and applies only if the density stratification is symmetric about the mid-channel, and under the Boussinesq approximation (these conditions are referred to as a \emph{symmetric fluid} below). Mode-2 ESWs of this type are henceforth denoted \textit{symmetric embedded solitary waves} (S-ESWs). Mode-2 ISW Euler computations without these symmetries have thus far yielded only GSWs \citep{vanden1992long,rusaas2002solitary}. Hence numerical evidence for generic mode-2 ``true" solitary waves was lacking, but recent laboratory observations point to their existence \citep{gavrilov2013influence,liapidevskii2018large}.

In a recent paper, \cite{barros2020strongly} found mode-2 ESW solutions to the MCC3 equations for `non-symmetric' fluids. In addition they found that these waves could have intricate `multi-hump' structures.
The goal of this paper is to present numerical evidence that mode-2 ESWs exist for the full Euler system beyond the idealised \emph{symmetric three-layer fluid}, together with elucidating the parameters for which they occur, their limiting configurations, and how they compare to those of MCC3. The bifurcation structure and wave profiles we find cannot be described by weakly nonlinear theories describing mode-2 ISWs such as the KdV equation.
We will find throughout that MCC3 acts as a useful guide to finding solutions to the full Euler equations.

Conjugate states of the system are central to understanding some of the limiting configurations we find. In both layered and continuously stratified fluid systems, mode-1 ISWs are oftentimes found to broaden as their amplitude increases, resulting in tabletop solitary waves, the limit of which being a wavefront connecting conjugate states (for example, this was rigorously demonstrated for a two-layer flow by \cite{amick1986global}). Conjugate states for a three-layer fluid system has been explored by \cite{lamb2000conjugate} and \cite{rusaas2002solitary}.    \cite{lamb2000conjugate} also compared three-layer conjugate states against those of a fluid with finite thickness pycnoclines, and found that the three-layer model is consistent with the continuously stratified model as the pycnocline thickness is decreased.

A favourable feature of the $\MCC$ equations is that they share the same conjugate states as the full Euler system, and we exploit this extensively. We find that 
 the $\MCC$ equations are an extremely powerful tool in qualitatively describing large amplitude solutions to the Euler equations, even in parameter regimes where the shallow water assumptions of the model are invalidated.

For concave mode-2 solitary waves, it was found recently for the $\MCC$ equations that the speed of the wave could exceed the mode-1 long wave speed \citep{barros2020strongly}. This is predicted by the observation that mode-2 conjugate states can have speeds greater than the linear long wave speed of mode-1. We show herein that concave mode-2 solitary wave branches outside the linear spectrum are also a feature of the full Euler system.

The paper is organised as follows. In section two, we formulate the problem to be solved, and introduce the steady $\MCC$ equations for comparison.  In section three, we explore the conjugate state equations.  In section four, we present and describe the numerical results. Concluding remarks are given in section five. In the Appendix, we present the numerical scheme used to compute fully nonlinear solutions.

\section{Formulation}\label{section:form}
\begin{figure}
	\begin{tikzpicture}
	\newcommand{\Hzero}{0.055}
	\newcommand{\Hone}{0.37}
	\newcommand{\Htwo}{0.47}
	\newcommand{\Hthree}{0.942}
	\newcommand{\xzero}{0.51}
	\node[anchor=south west,inner sep=0] (image) at (0,0) {\includegraphics[scale=1]{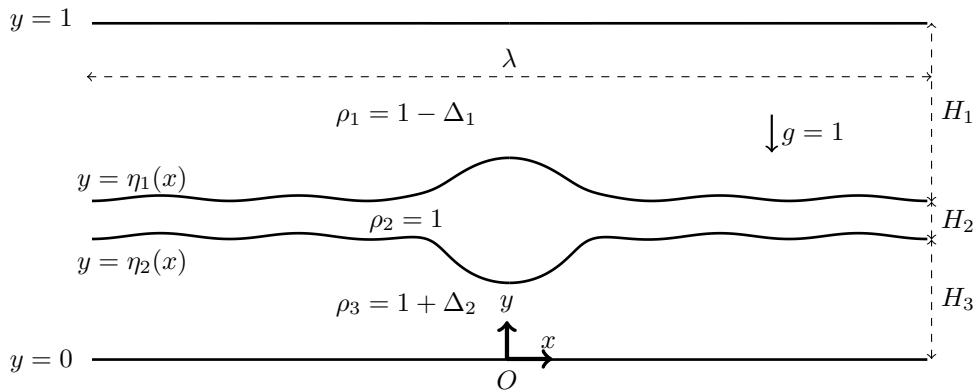}};
	\begin{scope}[x={(image.south east)},y={(image.north west)}]
	\draw [ dashed, <->]  (0.975,\Hzero) -- node[right]{$H_3$} (0.975,\Hone);
	\draw [ dashed, <->]  (0.975,\Hone) -- node[right]{$H_2$} (0.975,\Htwo);
	\draw [ dashed, <->]  (0.975,\Htwo) -- node[right]{$H_1$} (0.975,\Hthree);
	\node [] () at (0.1,0.31) {$y=\eta_2(x)$};
	\node [] () at (0.1,0.53) {$y=\eta_1(x)$};
	\node [] () at (0.4,0.7) {$\rho_1=1-\Delta_1$};
	\node [] () at (0.4,0.2) {$\rho_3=1+\Delta_2$};
	\node [] () at (0.4,0.42) {$\rho_2=1$};
	\node [] () at (0,.95) {$y=1$};
	\node [] () at (0,0.05) {$y=0$};
	\draw [ultra thick, ->]  (\xzero,\Hzero) -- (\xzero,\Hzero+0.1);
	\draw [ultra thick, ->]  (\xzero,\Hzero) -- node[above right]{$x$} (\xzero+0.05,\Hzero);
	\node [] () at (\xzero,0.2) {$y$};
	\node [] () at (\xzero,0) {$O$};
	\draw [ thick, ->]  (0.8,0.7) -- node[right]{$g=1$} (0.8,0.6);
	\draw [dashed, <->]  (0.05,0.8) -- node[above]{$\lambda$} (0.975,0.8);
	\end{scope}
	\end{tikzpicture}
	\caption{A sketch of the flow configuration. Three layers of constant density fluid are bounded between walls at $y=0$ and $y=1$. The layers are separated by interfaces $y=\eta_i$. \label{fig:conf}}
\end{figure}

Consider three immiscible fluids stably stratified in a two-dimensional channel of height $H$. The fluids are assumed to be incompressible and of constant density, and we assume that the resulting flow is irrotational. The density of each fluid is taken to be $\rho_i$,  $i=1,2,3$ (from top to bottom), and we require $\rho_3>\rho_2>\rho_1$ for a stable stratification. Subscripts will be used in this way for other quantities throughout the paper. We consider a periodic domain of length $\lambda$. 
Although this paper concerns ESWs with a flat far-field, the ESWs found in this paper are particular cases of GSWs, which we approximate with long periodic waves. The period is verified to be sufficiently long to not have appreciable effects on the ESWs. 
We take Cartesian coordinates $(x,y)$, where $y$ is perpendicular to the solid boundaries. The wave is assumed to be symmetric about $x=0$, and we choose $y=0$ at the bottom wall. 
We denote by $H_i$ the depth of fluid layer $i$ at the end of the wavelength, such that $H=H_1+H_2+H_3$.
 Throughout the paper, we choose $H$ and $\rho_2$ as the characteristic length and density respectively. 
We scale velocities such that the acceleration due to gravity $g$ is also taken to be unity. We restrict our attention to weak density stratifications in which the density jumps across the three layers are comparable, that is $\Delta_1\approx \Delta_2 \ll 1$ where $\Delta_i=\rho_{i+1}-\rho_i$.
The flow configuration is shown in figure \ref{fig:conf}. 

\subsection{Euler equations}
Let the velocity field in each layer be given by $\bm{u_i}=(u_i,v_i)$. There exists a velocity potential $\phi_i$ and streamfunction $\psi_i$ in each layer, such that
\begin{equation}\label{eq:u}
u_i = \frac{\partial\phi_i}{\partial x} =\frac{\partial\psi_i}{\partial y}, \hspace{1cm}  v_i = \frac{\partial\phi_i}{\partial y} = -\frac{\partial\psi_i}{\partial x}.
\end{equation}
Hence in each layer, the flow is governed by 
\begin{equation}\label{eq:lap}
\nabla^2 \phi_i = 0.
\end{equation}
We consider travelling wave solutions moving to the right with constant speed $c$. We take a frame of reference travelling with the wave, such that the quiescent flow far from a solitary wave has a horizontal velocity of $c$ in each layer.
We parameterise the lower and upper interfaces as $y=\eta_2(x)$ and $y=\eta_1(x)$ respectively. We also denote the interface displacements as $\zeta_i$, which reads
\begin{align}
\zeta_2=\eta_2-H_3, && \zeta_1=\eta_1-H_2-H_3. 
\end{align}
 As well as kinematic boundary conditions, that is 
\begin{equation}\label{eq:kBC}
\bm{u}\cdot \bm{\hat{n}}=0, \hspace{1cm} \text{for} \,\,\, \begin{cases}
y=0, \\
y=\eta_2(x), \\
y=\eta_1(x), \\
y=1, 
\end{cases}
\end{equation}
we enforce continuity of pressure on the interfaces, which (making use of Bernoulli's equation) gives
\begin{equation}\label{eq:Bern}
\frac{1}{2} \left(\rho_{i+1} \lvert \nabla \phi_{i+1} \rvert^2 - \rho_{i} \lvert \nabla \phi_{i} \rvert^2  \right) + \Delta_i \eta_i = B_i, \hspace{0.2cm} \text{for} \,\,\, y=\eta_i(x) \,\,\, (i=1,2).
\end{equation}
In the above, $\bm{\hat{n}}$ refers to the unit normal of the curve 
and $B_i$ are unknown constants to be found as part of the solution. The Boussinesq approximation, where differences in density only affect the buoyancy term, is commonly applied to scenarios where the density stratification is weak ($(\rho_3-\rho_1)/\rho_2\ll 1$). This modifies the above boundary condition to give
\begin{equation}\label{eq:Bern_bous}
\frac{1}{2} \left(\lvert \nabla \phi_{i+1} \rvert^2 -  \lvert \nabla \phi_{i} \rvert^2  \right) + \Delta_i \eta_i = B_i, \hspace{0.2cm} \text{for} \,\,\, y=\eta_i(x) \,\,\, (i=1,2).
\end{equation}
All the solutions we seek are assumed to be symmetric about $x=0$. 
Furthermore, we enforce
\begin{align}
u_3 &= u_2 = -c,  && \text{at} \,\, x=\pm\frac{\lambda}{2}, \,\, y=\eta_2, \label{eq:farfield1} \\
u_2 &= u_1 = -c,  && \text{at} \,\, x=\pm\frac{\lambda}{2}, \,\, y=\eta_1. \label{eq:farfield2}
\end{align}
These boundary conditions ensure that  (assuming the solution has no oscillatory tail) the wave propagates with constant speed $c$ to the right, in a frame of reference to which the wave is steady.

 For the numerical code, which is designed to compute periodic solutions, we introduce the variable  $\tilde{c}$, which is the wavelength averaged horizontal velocity in the bottom fluid for a fixed value of $y$:
\begin{equation}\label{eq:c1}
\tilde{c}=-\frac{1}{\lambda} \int\displaylimits_{-\lambda/2}^{\lambda/2} u_3(x,y) \, \mathrm{d} x ,\hspace{0.4cm} y=\text{constant}.
\end{equation}
The irrotationality of the flow ensures that the above value is independent of the value of $\hat{y}$ chosen. We will approximate solitary waves with long periodic waves. It is the case that, given the solitary waves are not generalised (i.e. the solution is flat in the far-field), as $\lambda\rightarrow \infty$, one finds $\tilde{c}\rightarrow c$.

\begin{figure}
\centering
\begin{overpic}[scale=1]{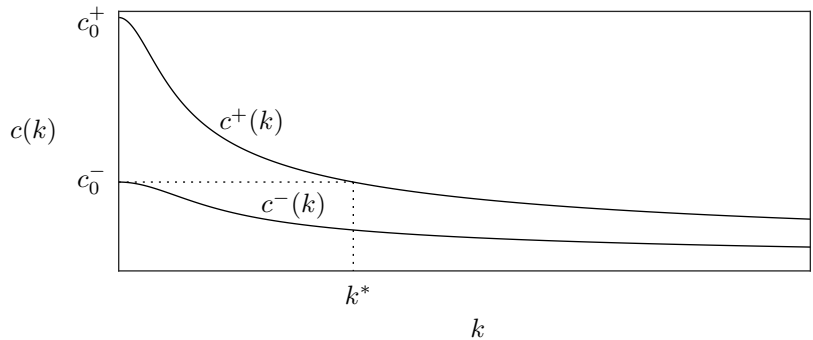}
\put(30,15){$c^-(k)$}
\put(25,25){$c^+(k)$}
\put(55,0){$k$}
\put(0,24){$c(k)$}
\put(40,4){$k^*$}
\put(8,37){$c_0^+$}
\put(8,18){$c_0^-$}
\end{overpic}
\caption{A typical dispersion relation, showing the fast mode ($c^+$) and slow mode ($c^-$) as functions of $k$. The value $k^*$ is such that $c^-(0)=c^+(k^*)$. \label{fig:disp}}
\end{figure}
Before describing model equations which approximate the above system, we first note that the above system has a linear dispersion relation \citep{rusaas2002solitary} given by 
\begin{align}\label{eq:disp_rel}
a_4 c^4 k^2 - a_2 c^2 k + a_0 = 0,
\end{align}
where $k$ is the wavenumber, and 
\begin{align}
a_4 &= 1 + \rho_3 C_2C_3 + \rho_3\rho_1 C_3C_1 + \rho_1C_1C_2, \\
a_2 &=  \Delta_2 \left( C_2 + \rho_1C_1\right) +  \Delta_1 \left( \rho_3 C_3 +  C_2\right),  \\
a_0 &= \Delta_1 \Delta_2,
\end{align}
and $C_i=\coth(H_i k)$. There exists two baroclinic modes: a faster mode (mode-1), denoted $c^+$, and a slower mode (mode-2), denoted by $c^-$. Both modes monotonically decrease from their maximum at $k=0$, denoted $c^\pm_0$.   Furthermore, it can be shown in the long-wave limit that the two interface displacements in the second (first) baroclinic mode have the opposite (same) polarities.
A typical dispersion relation is shown in figure \ref{fig:disp}. The resonant tail of a mode-2 GSW is associated with a mode-1 periodic wave which propagates at the same speed. Indeed, it can be seen in the figure that for mode-2 solitary waves bifurcating at $c_0^-$, there exists a finite value of $k=k^*$ such that $c_0^-=c^+(k^*)$.

We note that, under the Boussinesq approximation, the above system is invariant under the transformation $g\rightarrow -g$, and $\Delta_i\rightarrow -\Delta_i$. Furthermore, when the stratification and channel heights are symmetric ($H_1=H_3$ and $\Delta_1=\Delta_2$), we recover the \emph{symmetric three-layer fluid}. A feature of this parameter choice is the existence of families of mode-2 solutions with the property that $\zeta_1=-\zeta_2$. We refer to these as symmetric ESWs (S-ESWs).  The curve $y=0.5$ is a streamline, and the flow can be viewed as a two-layer flow with layer heights $H_3$ and $H_2/2$. Surprisingly, not all mode-2 solutions for the symmetric three-layer fluid have this additional symmetry, such as the large amplitude concave waves shown in section \ref{section:concave} (see figures \ref{fig:concave_branches}-\ref{fig:con_soln}). 

Fully nonlinear waves in the Euler equations are computed using the numerical method given in Appendix \ref{section:num_meth}. Below, we discuss the MCC3 model which approximates the three-layered full Euler system.

\subsection{Travelling waves for the $\MCC$ equations}\label{section:model}
We consider the three-layer Miyata-Choi-Camassa  equations (MCC3).
 We denote the shallow water parameter in each layer as $\epsilon_i=H_i/\lambda$, and the amplitude parameter as $\delta_i=\lvert\lvert\zeta_1\rvert\rvert_{\infty}/H_i$. The MCC3 model assumes $O(\epsilon_i)=O(\epsilon_j)$, has errors of $O(\epsilon_i^4)$, and has no smallness assumption on the parameter $\delta_i$. Hence, it is possible that the model can remain asymptotically valid for large amplitude solutions, provided the shallow water parameters in each layer are comparable and small. In a frame of reference travelling with a wave of constant speed $c$, and under the assumption that the flow in the far-field is unperturbed, one finds the Euler-Lagrange equations (see \citeauthor{barros2020strongly} \citeyear{barros2020strongly})
\begin{align}\label{eq:MCC_EL}
\frac{\partial L}{\partial \zeta_i} - \frac{d}{dx}\left[\frac{\partial L}{\partial \zeta_i'}\right] &=0, \hspace{1cm} i=1,2,
\end{align}
where the Lagrangian is given by
\begin{align}\label{eq:MCC_L}
L&= \frac{1}{6} c^2 \left[ \left(\rho_1 \frac{H_1^2}{h_1} + \frac{H_2^2}{h_2} \right) \zeta_{1}'^2  + \frac{H_2^2}{h_2} \zeta_{1}' \zeta_{2}'  + \left( \frac{H_2^2}{h_2} + \rho_3 \frac{H_3^2}{h_3} \right) \zeta_{2}'^2  \right] - V,
\end{align}
with a potential $V(\zeta_1,\zeta_2)$
\begin{align}\label{eq:MCC_V}
V &=  \frac{1}{2} \left\{ \left(\Delta_1 \zeta_1^2 + \Delta_2 \zeta_2^2 \right) - c^2 \left[ \left( \Delta_1 + \rho_1 \frac{H_1}{h_1} - \frac{H_2}{h_2} \right) \zeta_1  + \left( \Delta_2 - \rho_3 \frac{H_3}{h_3} + \frac{H_2}{h_2} \right) \zeta_2 \right] \right\}
\end{align}
In the above, $\zeta_i'=d\zeta_i/dx$.
The system is reduced, under the Boussinesq approximation, by removing every instance of $\rho_1$ and $\rho_3$, except where they appear in $\Delta_1$ and $\Delta_2$.

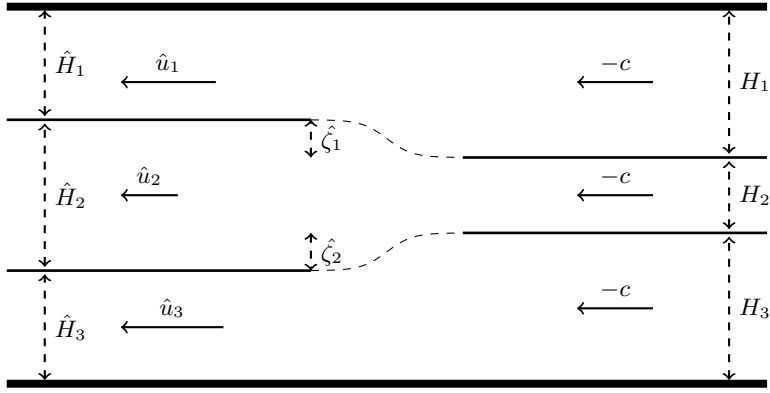
\begin{figure}
	\centering
	\begin{tikzpicture}
	
	\draw [ -,line width=3pt] (0,4) -- (10,4) ;
	\draw [ -,line width=3pt] (0,-1) -- (10,-1) ;
	\draw [ -,line width=1pt] (0,.5) -- (4,.5) ;
	\draw [ -,line width=1pt] (0,2.5) -- (4,2.5) ;
	\draw [ -,line width=1pt] (6,2) -- (10,2) ;
	\draw [ -,line width=1pt] (6,1) -- (10,1) ;
	
	\draw [ <->,thick]  (.5,-1+.05) -- (.5,.5-.05) [dashed] node [midway,right] {\small{$\hat{H}_3$}};
	\draw [ <->,thick]  (.5,.5+.05) -- (.5,2.5-.05) [dashed] node [midway,right]  {\small{$\hat{H}_2$}};
	\draw [ <->,thick]  (.5,2.5+.05) -- (.5,4-.05) [dashed] node [midway,right]  {\small{$\hat{H}_1$}};
	
	\draw [ <->,thick]  (4,2) -- (4,2.5) [dashed] node [midway,right]  {\small{$\hat{\zeta_1}$}};	
	\draw [ <->,thick]  (4,1) -- (4,.5) [dashed] node [midway,right]  {\small{$\hat{\zeta_2}$}};	
	
	\draw [ <->,thick]  (9.5,-1+.05) -- (9.5,1-.05) [dashed] node [midway,right] {\small{$H_3$}};
	\draw [ <->,thick]  (9.5,1+.05) -- (9.5,2-.05) [dashed] node [midway,right]  {\small{$H_2$}};
	\draw [ <->,thick]  (9.5,2+.05) -- (9.5,4-.05) [dashed] node [midway,right]  {\small{$H_1$}};
	\draw [ ->,thick]  (8.5,3)  -- (7.5,3)  node [midway,above]  {\small{$-c$}};
	\draw [ ->,thick]  (8.5,1.5) -- (7.5,1.5)  node [midway,above]  {\small{$-c$}};
	\draw [ ->,thick]  (8.5,0)  -- (7.5,0)  node [midway,above]  {\small{$-c$}};
	\draw [ ->,thick]  (2.75,3)  -- (1.5,3)  node [midway,above]  {\small{$\hat{u}_1$}};
	\draw [ ->,thick]  (2.25,1.5)  -- (1.5,1.5)  node [midway,above]  {\small{$\hat{u}_2$}};
	\draw [ ->,thick]   (2.85,-0.25) -- (1.5,-0.25)  node [midway,above]  {\small{$\hat{u}_3$}};
	
	\draw[domain=4:6, smooth, variable=\x, black, dashed] plot ({\x}, {2.25-.25*tanh(3.0*(\x-5)}); 
	\draw[domain=4:6, smooth, variable=\x, black, dashed] plot ({\x}, {0.75+.25*tanh(3.0*(\x-5)}); 
	\end{tikzpicture}
	\vspace{0.05cm}
	
	\caption{\label{fig:CS} Steady flow in a frame moving with a right-moving front.}
\end{figure}

Travelling wave solutions to \eqref{eq:MCC_EL} were the object of study in \cite{barros2020strongly}. Although GSWs were found to be prevalent, ESWs were numerically computed for special values of parameters. 
 Due to the additional condition imposed that the tails have no oscillations, the ESWs exist in a parameter space with one less dimension than that of GSWs.
The asymptotic limit $H_2\rightarrow 0$  was used to find `compacton' multi-hump solutions. These solutions, characterised by $p$ humps on the upper interface and $q$ on the lower,
 can be numerically extended into finite values of $H_2$ via the method of continuation. By doing so, mode-2 multi-hump GSWs were also exhibited,  and ESWs were found along these GSW branches.
Interestingly, it was suggested that not all truly localised mode-2 solitary waves are embedded, since concave solitary waves may be found outside the linear spectrum. These features will be further explored in the following sections. Here, computations of MCC3 solutions are performed with a pseudospectral collocation method.

\section{Conjugate states and limiting solitary waves of the $\MCC$ model}\label{section:form_CS}
 We assume the set-up illustrated in figure \ref{fig:CS} for an internal wavefront moving from left to right at constant speed $c>0$ into a three-layer stratified fluid at rest at $x=+\infty$. In a frame moving with speed $c$, the velocity at $x=+\infty$ is $-c$ in the three fluids.  
The depth of each layer at $x\rightarrow +\infty$ is given by $H_i$. The flow at $x\rightarrow-\infty$ has the lower and upper interface perturbed by $\hat{\zeta}_2$ and $\hat{\zeta}_1$ respectively.  We denote the new layer depths as $\hat{H}_i$ and the corresponding velocities as $\hat{u}_i$. The two horizontally uniform flows are said to be conjugate if all three basic physical conservation laws of mass, momentum and energy (for Euler equations) hold.

 Mass conservation in each fluid implies 
\begin{align}\label{eq:CS_u}
\hat{u}_i &= -\frac{cH_i}{\hat{H}_i}, && i=1,2,3,
\end{align}
where $\hat{H}_3 = H_3 + \hat{\zeta}_2$, $\hat{H}_2 = H_2 - \hat{\zeta}_2 + \hat{\zeta}_1$, and  $\hat{H}_1 = H_1 - \hat{\zeta}_1$.
There are three unknowns: $\hat{\zeta}_1$, $\hat{\zeta}_2$, and $c$. One equation comes from enforcing conservation of momentum, given by
\begin{equation}\label{eq:CS_mom}
\int_{x\rightarrow-\infty} P + \rho u^2 \, \mathrm{d}y = \int_{x\rightarrow +\infty} P + \rho u^2 \, \mathrm{d}y.
\end{equation}
Making use of Bernoulli's equation to find the pressure $P$,  \cite{lamb2000conjugate} showed this condition is equivalent to
\begin{align}\label{eq:momcon}
c^2 \left\{ (1+\Delta_2)\frac{\hat{\zeta}_2^3}{\hat{H_3}^2} + \frac{(\hat{\zeta}_1-\hat{\zeta}_2)^3}{\hat{H_2}^2} - (1-\Delta_1)\frac{\hat{\zeta}_1^3}{\hat{H_1}^2} \right\} &=0.
\end{align}
  The two equations which close the system are continuity of pressure across both interfaces. This is found to be
 \begin{align}
\frac{c^2}{2}\left[(1+\Delta_2) \lrr{1-\frac{H_3^2}{\hat{H_3}^2}} - \lrr{1-\frac{H_2^2}{\hat{H_2}^2}}\right] - \Delta_2 \hat{\zeta_2} &=0, \label{eq:cont_p2}\\
\frac{c^2}{2}\left[\lrr{1-\frac{H_2^2}{\hat{H_2}^2}} - (1-\Delta_1) \lrr{1-\frac{H_1^2}{\hat{H_1}^2}}\right] - \Delta_1 \hat{\zeta_1} &=0. \label{eq:cont_p1}
 \end{align}
These equations are explored by \cite{lamb2000conjugate} and \cite{rusaas2002solitary}. 
\cite{lamb2000conjugate} finds solutions by combining a  Newton-Ralphson iteration procedure with a a geometrical approach in which the conjugate states are found as intersection points of two plane algebraic curves.

\begin{figure}
    \centering
    \begin{overpic}[]{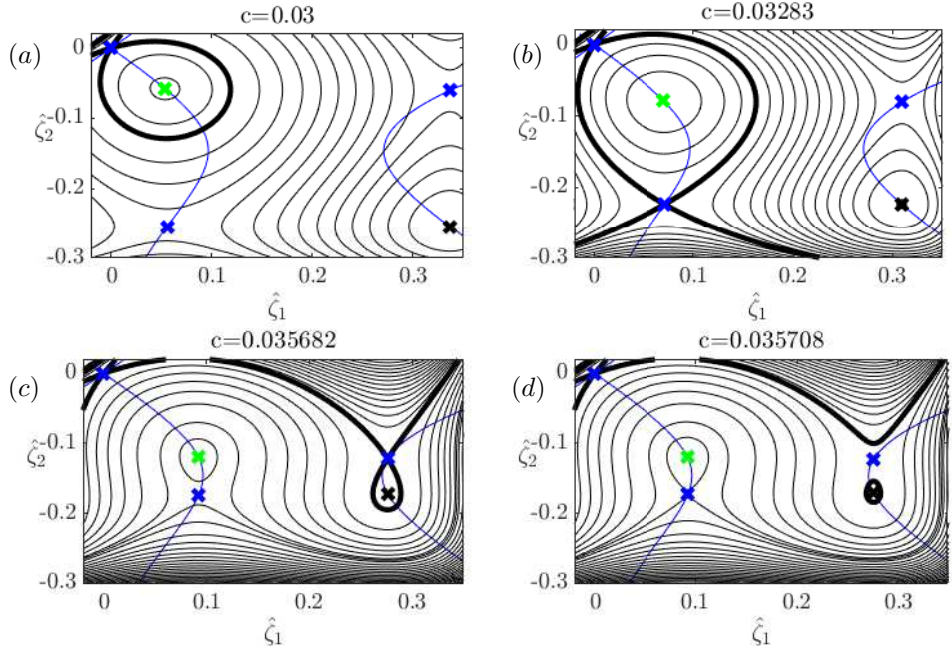}
    \put(3,58){$(a)$}
    \put(3,26){$(c)$}
    \put(51,58){$(b)$}
    \put(51,26){$(d)$}
    \end{overpic}
    \caption{Contour plot of the potential $V$ in the lower-right quadrant for given values of the speed $c\in(c_0^-,c_0^+)$. The thick black curve is the contour $V=0$. The critical points are given by the crosses, which are restricted to moving along the blue curves. The parameters are $H_2=0.03$, $H_1=1.2H_3$, and $\Delta_1=\Delta_2=0.01$. Excluding the origin, there are four critical points in the lower right quadrant: a local minimum, a local maximum, and two saddles, given by  green, black, and blue crosses respectively. The same colour scheme will be used throughout this section.}    
    \label{fig:crit_point}
\end{figure}

The conjugate state equations described above are recovered using the full Euler system.
 However, it can be seen immediately that equations \eqref{eq:cont_p2} and \eqref{eq:cont_p1} are equivalent to $V_{\zeta_1}=0$ and $V_{\zeta_2}=0$ respectively. Furthermore, it can be shown that equation \eqref{eq:momcon} is equivalent to
\begin{align}
2V - V_{\zeta_1} - V_{\zeta_2}=0.
\end{align} 
 Therefore, conjugate states in Euler equations can be viewed as non-trivial critical points of the potential of the $\MCC$ equations for which $V=0$. We remark that the ability of the strongly nonlinear theory to capture the same conjugate states for the fully nonlinear (Euler) theory has previosuly been shown for two-layer fluids with, or without, a top rigid lid by \cite{choi1999fully} and \cite{barros2016remarks}, respectively.

 At the origin $(\hat{\zeta}_1,\hat{\zeta}_2)=(0,0)$, the potential $V$ is equal to zero, and is a local minimum when $c\in(0,c_0^-)$, a saddle when $c\in(c_0^-,c_0^+)$, and a local maximum for $c>c_0^+$. For fixed $H_i$ and $\rho_i$, one can plot the critical points of $V$ for a given $c$, and observe how the critical points evolve as $c$ varies. As an illustration, consider figure \ref{fig:crit_point}, where we choose parameters $\Delta_1=\Delta_2=0.01$, $H_2=0.03$, and $H_1=1.2 H_3$. We restrict our attention to mode-2 conjugate states by considering the lower-right quadrant ($\hat{\zeta}_1>0$, $\hat{\zeta}_2<0$). For the given parameter values, $c_0^-=0.012$ and $c_0^+=0.069$. Therefore, $c\in(c_0^-,c_0^+)$ for all the contour plots in the figure, and hence the origin is a saddle of $V$. The black curves are level sets of $V$, where the bold curve corresponds to $V=0$. The blue curves, which are independent of $c$, are found by rearranging equations \eqref{eq:cont_p2}-\eqref{eq:cont_p1} to eliminate $c$, and are given by
\begin{align}
\Delta_1 \zeta_1\left[\rho_3 \lrr{1-\frac{H_3^2}{\hat{H_3}^2}} - \lrr{1-\frac{H_2^2}{\hat{H_2}^2}}\right] - \Delta_2 \zeta_2 \left[\lrr{1-\frac{H_2^2}{\hat{H_2}^2}} - \rho_1 \lrr{1-\frac{H_1^2}{\hat{H_1}^2}}\right] &=0.
\end{align}
The crosses correspond to critical points which must lie on the blue curves, and it can be seen there are four non-trivial such critical points in the lower-right quadrant. 
The critical points are a local minimum, two saddles, and a local maximum. These correspond with green, blue, and black crosses respectively. There are two further maxima, but these are outside the domain window. 
In panel $(b)$, the $V=0$ curve intersects one of the saddles, and the other saddle in panel $(c)$. Hence, the speeds $c$, and the values of $\hat{\zeta}_i$ at these saddles are conjugate states of the system. In panel $(d)$, the small contour enclosing the maximum will eventually (as $c$ increases) collapse to that point,  yielding a third mode-2 conjugate state in the lower-right quadrant.

\begin{figure}
	\centering
     \begin{overpic}[scale=1]{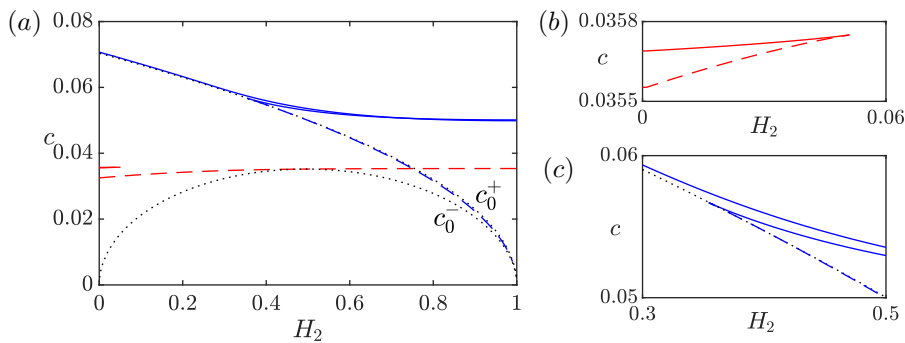}
     \put(0,35){$(a)$}
     \put(57.5,35){$(b)$}
     \put(58,19){$(c)$}
     \end{overpic}
     \caption{Solutions to the conjugate state equations for $\Delta_1=\Delta_2=0.01$, $H_1=1.2H_3$ with varying $H_2$. The black dotted curves are the linear long wave speeds. Red and blue curves are mode-2 and mode-1 conjugate states respectively. If the curve is dashed, then the conjugate state is a saddle of $V$, while solid curves are local maximums. Panels $(b)$ and $(c)$ are blow ups of different regions of the figure in panel $(a)$.}
     \label{fig:CS_op}
 \end{figure}

In the spirit of \cite{dias2001interfacial} we plot in figure \ref{fig:CS_op} branches of conjugate states for the parameter regime $\Delta_1=\Delta_2=0.01$ and $H_1=1.2 H_3$. The figure shows how the number and speed of conjugate state solutions vary with increasing values of $H_2$. The black dotted curves are $c_0^-$ and $c_0^+$. 
 Blue curves are mode-1 conjugate states (sought in the first and third quadrants of the $(\zeta_1,\zeta_2)$-plane), while red curves are mode-2 (sought in the second and fourth quadrants of the $(\zeta_1,\zeta_2)$-plane). If the curve is solid, the conjugate state is a maximum of $V$, while dashed curves denote a saddle.

Let us first discuss mode-1 conjugate states. It can be seen that for smaller values of $H_2$ there is one maximum conjugate state with speeds $c>c_0^+$, while for larger values of $H_2$ there are two. In both cases, our numerical tests show that heteroclinic orbits from the maximum origin ($c>c_0^+)$ to a maximum conjugate state can be found. Hence, when $H_2$ is sufficiently large, two heteroclinic orbits are found. Furthermore, they exhibit different polarities.
 There is also a saddle mode-1 conjugate state with speeds slightly below $c_0^+$. However, no heteroclinic orbits or large amplitude solutions related to this critical point were found numerically,  in full agreement with the results by \cite{lamb2000conjugate}.

 \begin{figure}
 \centering
 \begin{overpic}[scale=1]{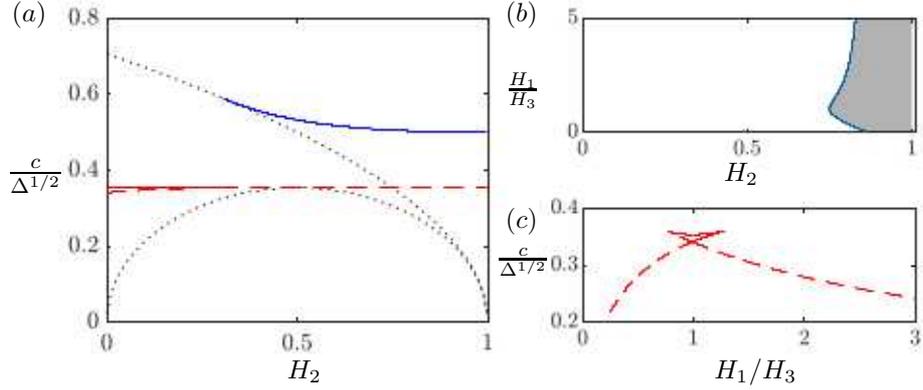}
 \put(0,38){$(a)$}
 \put(-.5,21){\large $\frac{c}{\Delta^{1/2}}$}
 \put(29,0){$H_2$}
 \put(52,38){$(b)$}
 \put(52,30){$\frac{H_1}{H_3}$}
 \put(75,21){$H_2$}
 \put(52,16){$(c)$}
 \put(51,12){$\frac{c}{\Delta^{1/2}}$}
 \put(74,0){$H_1/H_3$}  
 \end{overpic}
 \caption{All the panels concern conjugate states with the Boussinesq assumption and $\Delta_1=\Delta_2$. Panel $(a)$ shows conjugate states for the symmetric three-layer fluid ($H_1=H_3$) with varying $H_2$, using the same colour and line scheme as in figure \ref{fig:CS_op}. In panel $(b)$, the parameter region where mode-2 conjugate states have speeds exceeding $c_0^+$ is shaded. In panel $(c)$, we plot the mode-2 conjugate states, with infinitesimal $H_2$, as a function of $H_1/H_3$. Dashed curves are saddles of the $\MCC$ potential $V$, while the solid curve is a maximum. }
 \label{fig:newfig_CS_bous}
\end{figure}

For mode-2, there are either three or one mode-2 conjugate states. While precise formulae separating the regions of parameter space for which three or one occur were not provided, \cite{lamb2000conjugate} demonstrated that two of the conjugate states seize to exist for $H_2$ above a critical value. This critical value decreases and ultimately becomes zero as one deviates further from a symmetric three-layer fluid, either by breaking the depth or stratification symmetry, or in Lamb's case by increasing the strength of the stratification while removing the Boussinesq approximation.  The parameters for figure~\ref{fig:CS_op} have three mode-2 conjugate states for $H_2<0.051$: one maximum and two saddles. For $H_2$ larger than this value, only a saddle remains. Therefore, the maximum is restricted to small values of $H_2$, and it lies within the linear spectrum of the system. There is a special value of $H_2=H_2^*$ at which the curve for the mode-2 conjugate state is tangent to the curve $c=c_0^-$.  It can be shown that this corresponds to the criticality condition for the KdV equation for which the quadratic nonlinearity coefficient vanishes. For the symmetric three-layer fluid, this is given precisely by $H_2^*=0.5$. The KdV theory predicts convex (concave) mode-2 waves for values of $H_2$ less (greater) than $H_2^*$. This characterisation for the polarity of solutions seems to be in agreement with numerical solutions found in this paper.

We note that when the Boussinesq approximation is considered, similar considerations apply. 
However, for a symmetric three-layer fluid, 
the criticality condition of the  KdV equation for mode-1 is always met regardless of the value of $H_2$. As a consequence, the mode-1 KdV equation has no  solitary wave solutions. When higher order nonlinearities are accounted for, mode-1 solutions are expected to exist provided $H_2$ is sufficiently large. This is confirmed for the $\MCC$ model, and it can be shown that the curve for the mode-1 conjugate state starts precisely at $H_2=4/13$, in complete agreement with Gardner theory  \citep{talipova1999cubic,lamb2006extreme}. Figure \ref{fig:newfig_CS_bous}(a) shows the conjugate states for this configuration. While the figure appears to only show one mode-1 conjugate state, the blue curve corresponds to two different conjugate states, under the symmetry $\Delta_i\rightarrow-\Delta_i$, $g\rightarrow -g$.
    
We would like to emphasise a novel aspect of mode-2 conjugate state solutions. As the value of $H_2$ increases beyond $H_2^*$, it may be that a mode-2 saddle conjugate state exceeds $c_0^+$, in which case mode-2 solitary wave solutions are no longer embedded. To unveil the regions in parameter space for which the mode-2 conjugate state exits the linear spectrum, we include the Boussinesq assumption and set  $\Delta_1=\Delta_2$. The results are shown in figure 
\ref{fig:newfig_CS_bous}$(b)$ and, as expected, such a feature only occurs for thick intermediate layers. When $H_1=H_3$, the value of $H_2$ for which this occurs is  $H_2=6H_1$ (resulting in $H_2=0.75$) in agreement with figure \ref{fig:newfig_CS_bous}$(a)$. In the case when $H_2$ is small, we can use asymptotics to predict for which parameters there is one (or three) mode-2 conjugate states, as shown in figure \ref{fig:newfig_CS_bous}$(c)$. In agreement with \cite{lamb2000conjugate}, the three conjugate states exist when $H_1/H_3$ is close to unity.

\begin{figure}
    \centering
    \begin{overpic}[scale=1]{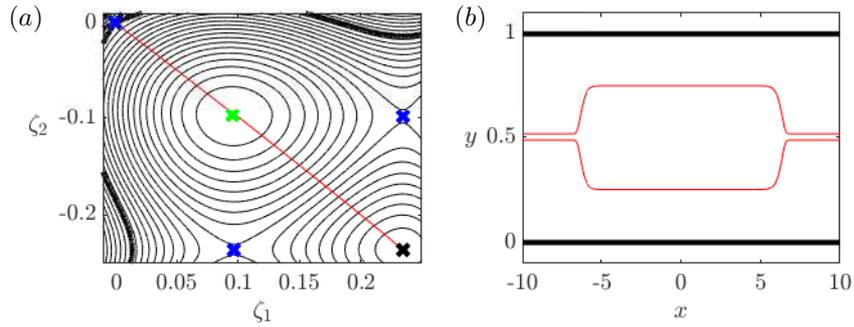}
    \put(3,32){$(a)$}
    \put(50,32){$(b)$}
    \end{overpic}
    \caption{A mode-2 $\MCC$ convex ESW with $\Delta_1=\Delta_2=0.01$, $H_2=0.03$, $H_1=0.9891H_3$, and $c=0.0354$ ($c_0^-<c<c_0^+$). In panel $(a)$, we show the projection of the solution (in red) over the $(\zeta_1,\zeta_2)$-plane. Panel $(b)$ shows the solutions in the physical space. This ESW is close to a heteroclinic orbit from the origin to a maximum of the potential $V$. }
    \label{fig:saddlemaxnew}
\end{figure}
\begin{figure}
    \centering
    \begin{overpic}[scale=1]{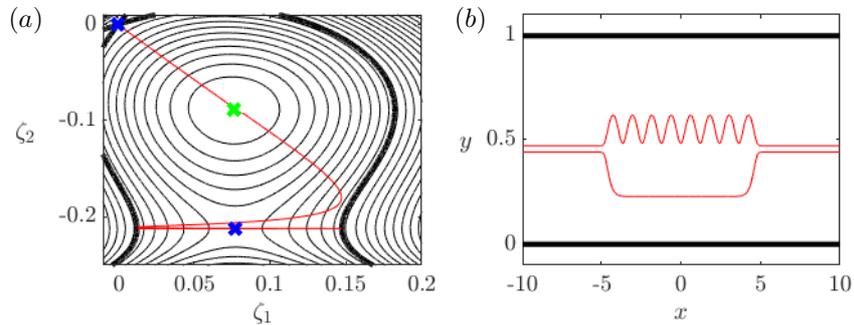}
    \put(3,32){$(a)$}
    \put(50,32){$(b)$}
    \end{overpic}
        \caption{A mode-2 $\MCC$ convex ESW wave with $\Delta_1=\Delta_2=0.01$, $H_2=0.03$, $H_1=1.2H_3$, and $c=0.03372$  ($c_0^-<c<c_0^+$). In panel $(a)$, we show the projection of the solution (in red) over the $(\zeta_1,\zeta_2)$-plane.  Panel $(b)$ shows the solutions in the physical space. Solutions of this type necessarily require speeds greater than the speed such that the saddle is a conjugate state of the system, since we require the value of $V$ at the saddle to be less than zero. }
    \label{fig:saddlesaddlenew}
\end{figure}
\begin{figure}
    \centering
    \begin{overpic}[scale=1]{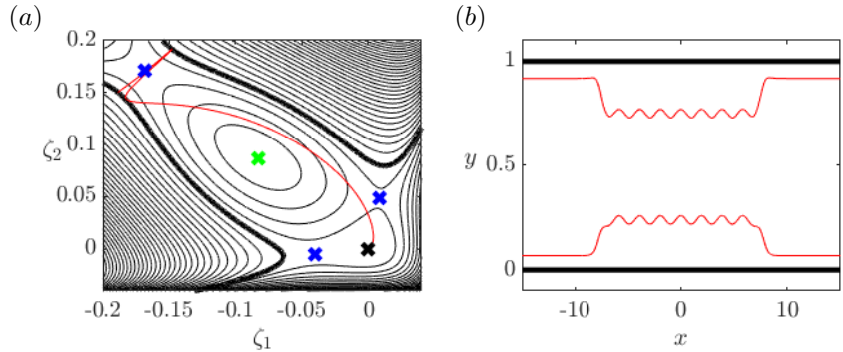}
    \put(3,35){$(a)$}
    \put(50,35){$(b)$}
    \end{overpic}
    \caption{A mode-2 $\MCC$ concave solitary wave with $\Delta_1=\Delta_2=0.01$, $H_2=0.85$, $H_1=1.2H_3$, and $c=0.0359$ ($c>c_0^+$). In panel $(a)$, we show the projection of the solution (in red) over the $(\zeta_1,\zeta_2)$-plane. Panel $(b)$ shows the solutions in the physical space. Solutions of this type necessarily require speeds greater than the speed of the saddle conjugate state of the system, since we require the value of $V$ at the saddle about which the solution oscillates to be less than zero. }
    \label{fig:maxsaddlenew}
\end{figure}

\subsection{Limiting solitary waves for MCC3}
The existence of a solution to the conjugate state equations does not imply the existence of a connection between conjugate states within the fluid equations. Below, we explore whether or not connecting orbits to equilibrium exist for the conjugate states in the context of the MCC3 system. This is done by seeking large amplitude solitary waves, where a broad tabletop solitary wave is numerical evidence that heteroclinic connections exist.
The Hamiltonian structure of the MCC3 equations allows deeper insight in to the solution behaviour.  We note beforehand that much of this behaviour persists for the Euler system, even in parameter regimes where the MCC3 system performs poor quantitatively, as seen in section 4. Retaining the same conjugate states as the full Euler system is a key strength of the $\MCC$ model.

First, we seek heteroclinic orbits from the origin to a mode-2 conjugate state which is a maximum of the MCC3 potential $V$. As seen in the previous section, these conjugate states are restricted to small $H_2$ and must lie within $c_0^-<c<c_0^+$, and hence the origin is a saddle. Numerical results show we are able to find heteroclinic orbits a subset of these conjugate states. Since the speeds are within the linear spectrum, solitary waves typically have oscillatory tails (GSWs), and parameter values must be chosen carefully to ensure the oscillations are of zero amplitude (how to find such parameters is discussed in section 4). As an example,  figure \ref{fig:saddlemaxnew} shows a mode-2 tabletop ESW for $\Delta_1=\Delta_2=0.01$ and $H_2=0.03$, and $H_1=0.9891H_3$.  Clearly, this large-amplitude solution is close to a flat wavefront connecting the origin to a maximum of $V$.

Next, consider saddle mode-2 conjugate states. These were found to exist for all values of $H_2\neq H_2^*$, and for sufficiently large $H_2$ they have speeds $c>c_0^+$. When the conjugate state has a speed $c_0^-<c<c_0^+$, we were unable to find any tabletop solitary waves (except for the special S-ESW solution branch). 
However, interesting multi-hump solitary waves were found. They are characterised by oscillations along the broadened section of the wave and with speeds exceeding that of the conjugate state, as in figure \ref{fig:saddlesaddlenew}. This result is surprising as it demonstrates that the speeds of mode-2 saddle conjugate states may not be limiting speeds for mode-2 solitary waves. Parameters must again be chosen carefully to avoid oscillatory tails. For saddle mode-2 conjugate states with speeds $c>c_0^+$, large amplitude solitary waves are again typically characterised by oscillations on the broadened section, as in figure \ref{fig:maxsaddlenew}. These solutions differ from those with speeds $c_0^-<c<c_0^+$ since they form a family of continuous solutions in speed-amplitude bifurcation space for fixed $H_i$ and $\Delta_i$. Furthermore, special parameter values can be found such that the oscillations on the broadened section vanish, resulting in a tabletop solitary wave.

To summarise, large amplitude mode-2 solitary waves for the MCC3 system are related to the conjugate states of the Euler equations. If the conjugate state has a speed $c_0^-<c<c_0^+$, the origin is a saddle and careful selection of parameters must be chosen to find ESWs. For weak stratifications, if the conjugate state is a maximum, limiting solutions (if they exist) are tabletop solitary waves. Meanwhile, if the conjugate state is a saddle, multi-hump solitary waves are found with speeds faster than that of the conjugate state. Except for the symmetric Boussinesq solutions, we were only able to find tabletop solitons predicted by a saddle conjugate state when $c>c_0^+$. This does not mean such solutions do not exist, but the parameter space in which they exist is very restricted, requiring the suppression of oscillations at both the origin and the conjugate state.

Large amplitude solutions to the full Euler system are explored further in section \ref{section:results}, where solutions with the characteristics seen in  figure \ref{fig:saddlemaxnew}--\ref{fig:maxsaddlenew} are found.

\section{Results for the Euler equations and comparison with MCC3}\label{section:results}   
In this section, we discuss numerically computed mode-2 solitary waves for the three-layer configuration. 
 Comparisons between the fully nonlinear (Euler) and strongly  nonlinear (MCC3) theories are made, and the conjugate states discussed in section \ref{section:form_CS} are referred to when describing large amplitude solutions.

 \begin{figure}
	\centering
	\begin{overpic}[scale=1]{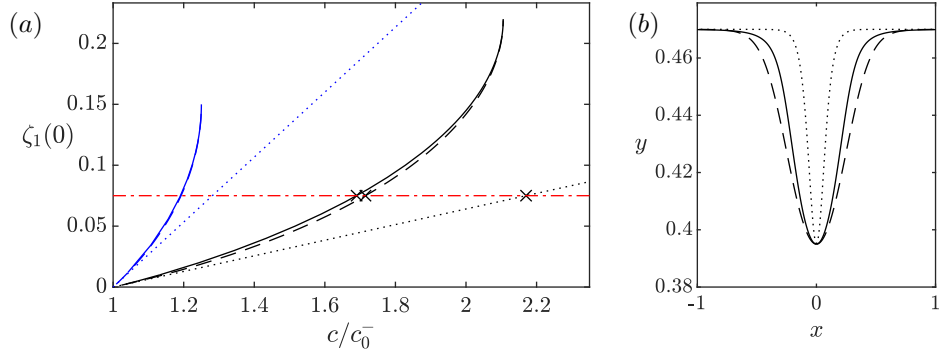}
	\put(0,35){$(a)$}
	\put(65,35){$(b)$}
	\end{overpic}	
	\caption{Panel $(a)$ shows 
	S-ESW solution branches. The black and blue curves correspond to $H_2=0.06$ and $H_2=0.2$ respectively. The dashed, dotted, and solid curves correspond to the KdV, MCC2, and full Euler branches respectively. Each cross in panel $(a)$ is a solution which intersect the red dotted-dashed curve $\zeta_1(0)=0.075$. These are shown in panel $(b)$ using the same line-styles, where only the lower interface
is shown. The upper interface is a reflection of this profile about $y = 0.5$   \label{fig:newfig10}}
\end{figure}

\subsection{Convex waves}\label{section:convex}
We  first consider the case where parameters are chosen such that the waves are convex. From section \ref{section:form_CS}, this occurs for values of $H_2$ below a value $H_2^*$ given by the criticality condition. As such, the conjugate states are attained at speeds $c<c_0^+$. Hence, we expect mode-2 solitary waves to be within the linear spectrum, typically characterised by oscillatory tails. Along these branches of GSWs, special values of the parameters are found such that the tails have zero amplitude.
\subsubsection{Symmetric embedded solitary waves (S-ESWs)}\label{section:results_sym}

 Along branches of GSWs, the parameters for which the solutions have no oscillations in the far-field are typically  not known a priori, and must be found as part of the solution. This is not the case, however, for a symmetric three-layer fluid.  With this configuration, there exists for all $H_2\neq H_2^*=0.5$ a branch of mode-2 solitary waves bifurcating from zero amplitude at $c_0^-$ and ending in a heteroclinic connection between the origin and a conjugate state. 
  The interface displacements of these solutions are related via $\zeta_1=-\zeta_2$, and can be constructed by reflecting a two-layer mode-1 solution across an imaginary bounding wall.

An extensive comparison of KdV, MCC2, and Euler solitary waves in a two-fluid system can be found in \cite{camassa2006realm}. They found that the MCC2 model performs well up to the limiting tabletop solitary wave, given the shallow water assumptions hold. Interestingly,  they observe the largest deviation between the Euler and MCC2 model are for waves of moderate amplitude, rather than the largest limiting waves. We demonstrate this in figure \ref{fig:newfig10}(a), which shows symmetric Boussinesq solitary wave branches in the  $(c/c_0^-,\zeta_1(0))$-plane
 for $H_2=0.06$ and $H_2=0.2$. We have set $\Delta_1=\Delta_2$, where the value chosen does not matter, since for the Boussinesq system, any change can be expressed as a rescaling of $c$. In the figure, we plot the branches for the KdV model (dotted curves), the MCC2 model (dashed curves), and full Euler (solid curves). The black curves correspond to $H_2=0.06$, while the blue curves are for $H_2=0.2$, and the agreement can be seen to be better for the blue curves.
 This is due to a more severe invalidation of the shallow water assumption (i.e. that $\epsilon_3\sim\epsilon_2 \ll 1$) for $H_2=0.06$. For example, using the `effective wavelength' as the horizontal length scale (see \cite{koop1981investigation}) one finds that the Euler solutions with $\zeta_1(0)=0.075$ from figure \ref{fig:newfig10}$(a)$ have $\epsilon_3= 2.0369$ for $H_2=0.06$ and $\epsilon_3=0.73542$ for $H_2=0.2$. The lower interface of this solution with $H_2=0.06$ is shown in panel $(b)$ of figure \ref{fig:newfig10}, and it can be seen that the MCC2 model is not a great approximation of the Euler equations here. Despite this, the solution branches of the MCC2 system has the same behaviour as that of the Euler branch (as opposed to say the KdV branches), and in particular limits to a heteroclinic orbit to the same conjugate state as the Euler system. This implies that the MCC2 model retains the correct information for describing (at least qualitatively) large amplitude solutions in regions of parameter space one may expect it to perform poorly.

 Below, we discuss the solution space for convex mode-2 solitary waves to the MCC3 and full Euler models when this symmetry is broken. 

\subsubsection{Non-symmetric configuration}\label{section:results_symbrk}
\begin{figure}
    \centering
    \begin{overpic}[scale=1]{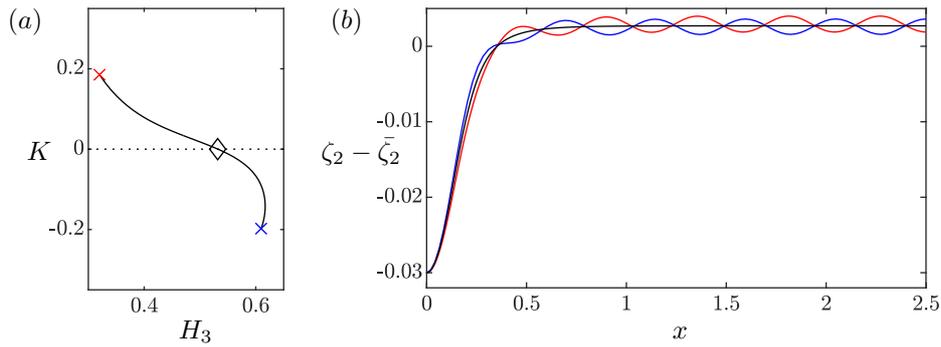}
    \put(0,35){$(a)$}
    \put(34,35){$(b)$}
    \put(33,21){$\zeta_2-\bar{\zeta_2}$}
    \put(2,21){$K$}
     \put(18,2){$H_3$}
    \put(70,2){$x$}
    \end{overpic}
    \caption{Panel $(a)$ shows a branch of GSW with $\Delta_1=\Delta_2=0.01$, $H_2=0.06$, and $\zeta_2(0)-\bar{\zeta_2}=-0.03$ (where $\bar{\zeta_2}$ is the mean of $\zeta_2$ over a wavelength) with varying values of $H_3$ (and $H_1$). The dotted curve is $K=0$. The red and blue profiles in panel $(b)$ correspond to the red and blue crosses in panel $(a)$, while the black profile is the $K=0$ solution (given by the diamond).  }
    \label{fig:newfig12}
\end{figure}
It is known that the oscillations in the tail of a GSW can arise from exponentially small terms \citep{sun1993exponentially,grimshaw1995weakly,sun1999non}. Therefore, one must take care in ensuring the solution computed is truly localised, as opposed to the resonant tail being of an order smaller than the capabilities of the numerical scheme.
 \cite{champneys2002true} proposed a criterion based on the continuity of curvature as a function of bifurcation parameters which we adopt here. Denoting the curvature of the lower interface at the last meshpoint in the tail as $K$, where $K$ is counted positive if the radius of curvature lies inside the bottom fluid, we state that an ESW is found along branches of GSW when the branch passes through $K=0$. The justification is as follows: the domain is periodic, and hence the solution ends on a wave trough if $K>0$, and a wave crest if $K<0$. The solution which occurs at $K=0$ must have waves with zero amplitude in the tail. This technique has been used to explore the existence of ESW for fully nonlinear gravity-capillary waves \citep{champneys2002true} and gravity-flexural waves \citep{gao2016nonlinear}. In all the cases above,  no ESWs were found. In fact, except for symmetric Boussinesq mode-2 waves, positive results for the existence of ESWs are only available for reduced long wave models (see e.g.  \cite{champneys2002true} for ESWs in a fifth-order modified KdV equation; \cite{barros2020strongly} for ESWs in the $\MCC$ model). The results to follow are the first positive result in the context of the full Euler equations.

Using an S-ESW with $H_3=0.47$, $H_2=0.06$, and $\Delta_1=\Delta_2=0.01$ as an initial guess, we break the symmetry by removing the Boussinesq approximation. The numerical scheme described in Appendix \ref{section:num_meth} then converges to a GSW. We fix the perturbation of the lower interface $\zeta_2(0)$ and vary the value of $H_3$ to obtain a branch of GSW, shown in figure \ref{fig:newfig12}$(a)$. We plot the solution branch with $K$ on the vertical axis, and it is observed that the solution branch passes through $K=0$ at the black diamond. 
Figure \ref{fig:newfig12}$(b)$ shows $\zeta_2$ for the three solutions along the branch. It demonstrates how $K$ going from positive to negative transitions through a localised solution.

\begin{figure}
	\centering
\begin{overpic}[scale=1]{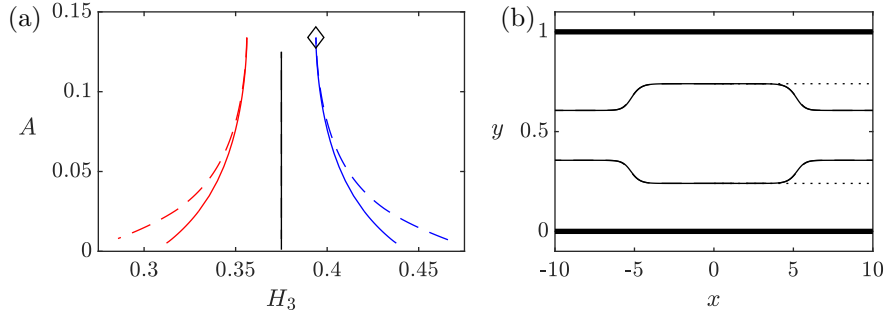}
\put(54,32){(b)}
\put(0,32){(a)}
\end{overpic}
	\caption{Panel $(a)$ shows Boussinesq ESW branches for the $\MCC$ (dashed curves) and Euler (solid curves), with fixed $H_2=0.25$. We vary the stratification, where the red curves have  $\Delta_1=0.01$, $\Delta_2=0.011$ while the blue curves have $\Delta_1=0.011$, $\Delta_2=0.01$ . The $\MCC$ and Euler S-ESW branches ($\Delta_1=\Delta_2$) overlap, given by the black curve.
 Panel $(b)$ shows the solution given by the diamond in panel $(a)$ for both the $\MCC$ (dotted curve) and Euler branch (dashed curve). 
	The profiles are almost indistinguishable. The dashed line shows a solution to the conjugate state equations given the incoming flow from the left. To four significant figures, both the Euler and $\MCC$ solution have the parameter values $H_3=0.3561$, $H_1=0.3939$, and $c =0.03618$.	\label{fig:newfig16}}
\end{figure}

We now modify the procedure to enforce the condition that $K=0$. As suggested by figure \ref{fig:newfig12}$(a)$, this requires removing a degree of freedom from the problem: for example, instead of fixing the amplitude and varying the speed (or vice versa), we must allow both to vary. Alternatively, one can allow an additional parameter to vary (for example, fix the amplitude and allow $c$ and $H_1$ to vary). The above procedure allows the computation of branches of ESWs. 
In the discussion that follows, we compute branches of ESWs by fixing $H_2$, $\Delta_1$, and $\Delta_2$, and varying the value of $H_3$. All the solutions in this subsection retain the bulge profile of S-ESWs. Much richer solutions will be presented in following sections.

\begin{figure}
	\centering
	\begin{overpic}[scale=1]{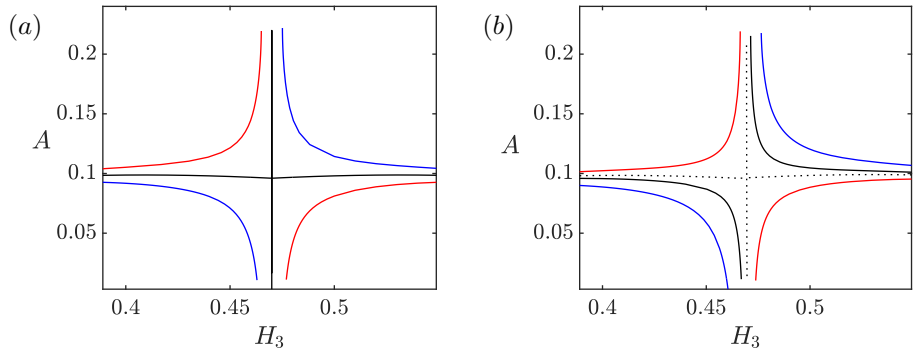}
	\put(0,35){$(a)$}
	\put(50,35){$(b)$}
	\end{overpic}
	\caption{Both panels show $\MCC$ ESW branches with fixed $H_2=0.06$ and $\Delta_1=0.01$. Panel $(a)$ is with the Boussinesq approximation, and panel $(b)$ is without. For both panels, the value of $\Delta_2$ is $\Delta_2=0.0098$, $\Delta_2=0.01$ and $\Delta_2=0.0102$ for the blue, black, and red curves respectively. The dashed black curve in (b) has $\Delta_2=0.0101575$. \label{fig:newfig21}}
\end{figure}

Let $A$ be defined by
\begin{equation}
A=\max \left(\lvert \zeta_1(0) \rvert,\lvert \zeta_2(0)\rvert\right).
\end{equation}
Figure \ref{fig:newfig16}$(a)$ shows the three branches of Boussinesq ESWs plotted in the $(H_3,A)$-plane with $H_2=0.25$, and $\Delta_2=1.1\Delta_1 $ (red curves), $\Delta_1=1.1\Delta_2$ (blue curves), and  $\Delta_1=\Delta_2$ (black curves). The dashed curves are $\MCC$ solutions, while the solid curves are Euler solutions. The black curves correspond to 
an S-ESW branch (i.e. $H_1=H_3$ and $\Delta_1=\Delta_2$). A break in the stratification symmetry results in a solution branch either side of the $\Delta_1=\Delta_2$ branch. As $A$ increases along the branch, the waves get broader, becoming a tabletop solitary wave, and limiting to a wavefront.  
A tabletop solitary wave for the blue branch is shown in panel $(b)$, where a solution of the conjugate state equations (which is a maximum of the $\MCC$ potential $V$) is given by the dotted curves.
It is not clear from the numerical scheme how the solution branches terminate at the other end. The Euler and $\MCC$ numerical solvers failed to converge beyond the points plotted in the figure. The amplitude of the waves by this point is very small.

The behaviour described above differs greatly from weakly nonlinear KdV or Gardner theory,  which predicts full branches of mode-2 localised solitary waves for given $H_i$ and $\Delta_i$. Meanwhile, the MCC3 system captures the reduction of dimension of the parameter space for these ESWs: given $H_i$ and $\Delta_i$, localised solitary waves of mode-2 are found as isolated points on branches of GSWs (except for the heavily idealised S-ESW solutions).

Keeping other parameters the same, but removing the Boussinesq approximation, one finds a similar solution space. The key difference is that, unlike the Boussinesq case, the branch with symmetric density stratification does not lie on the curve $H_1=H_3$, and the solutions are not symmetric about $y=0.5$.

As $H_2$ is decreased, features seen for the $H_2=0.25$ solution space persist in the full Euler equations. However, we found that the code failed to converge for very large amplitudes when $H_2$ was too small ($H_2<0.06)$, and hence we were not able to approach a limiting wavefront solution. While reasonable agreement is found between the Euler and $\MCC$ ESW branches when $H_2=0.25$, for smaller $H_2$ stark differences appear.
We suspect the cause of the increased discrepancy is the invalidation of the shallow water approximation, which resulted in worse agreement for the symmetric Boussinsq solutions in section \ref{section:results_sym}. While the agreement for small $H_2$ is poor with the Euler system, we describe below the behaviour of the MCC3 solution branches.

 Figure \ref{fig:newfig21} shows branches of ESWs for the $\MCC$ equation with and without the Boussinesq approximation in panel $(a)$ and $(b)$. We fix $\Delta_1=0.01$, $H_2=0.06$, and vary $\Delta_2$. For the Boussinesq case, when $\Delta_2=\Delta_1$, the ESW branch with $H_1=H_3$ is recovered, as above. However, along this branch there is a bifurcation point occurring about $A\approx 0.1$. The bifurcating branch no longer satisfies $H_1=H_3$. Numerical computations show that this bifurcation point occurs for smaller values of $A$ as $H_2$ is increased, and no longer exists for $H_2\approx 0.22$. Breaking the density symmetry results in the red and blue curves, which lie in domains bounded by the $\Delta_1=\Delta_2$ curves. The branches for large $A$ ultimately end in a wavefront, while again it is unclear from the numerical results how the branches ends with small values of $A$ terminate. For the non-Boussinesq model, we no longer have the symmetric Boussinesq solution  branch. Panel $(b)$ shows branches for four choices of $\Delta_2$. The $\Delta_2=\Delta_1$ branch without the Boussinesq approximation no longer lies on $H_3=H_1$. However, for a nonsymmetric stratification ($\Delta_2=0.0101575$, given by the dotted curve), a solution branch with a bifurcation point is found. Note that, in agreement with the results of \cite{camassa2006realm} for two-layer systems, the deviation from the Euler system is most pertinent at moderate amplitude. For large amplitude, limiting wavefront solutions are found for both the Euler and MCC3 system, and the interfaces and wave parameters are comparable.

\begin{figure}
	\centering
	\begin{overpic}[scale=1]{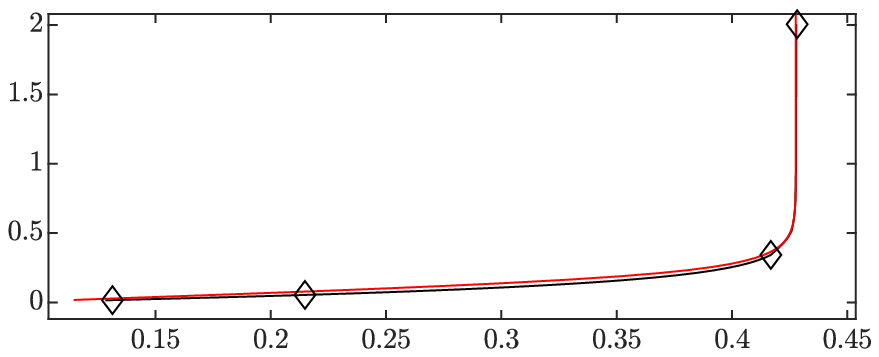}
	\put(50,0){$H_3$}
	\put(2,23){$Q$}
	\put(18,12){$(1)$}
	\put(35,13){$(2)$}
	\put(81,17){$(3)$}
	\put(81,34){$(4)$}
	\end{overpic}
	\caption{A multi-hump ESW branch with $\Delta_1=\Delta_2=0.01$ and $H_2=0.15$ for the $\MCC$ model (red) and Euler model (black). The solutions (1)--(4) along the $\MCC$ branch are shown in figure \ref{fig:H2_015_soln} by the red curves, while the Euler solution with the same $H_3$ is shown in black. \label{fig:H2_015_branch}}
\end{figure}

\subsubsection{Multi-hump solitary waves}\label{section:multi}
Thus far, the ESWs we have seen are the typical `bulge-shaped waves' observed in laboratory experiments (for example, \citeauthor{carr2015experiments} \citeyear{carr2015experiments}). \cite{barros2020strongly} found compacton solutions for the MCC3 system comprised of $p$ and $q$ humps on the lower and upper interface respectively (referred to as a $(p,q)$ compacton). They were recovered in the asymptotic limit $H_2\rightarrow 0$, but were found numerically to persist for finite values of $H_2$. Such solutions were coined multi-hump solitary waves, and appear to have been observed in the laboratory by \cite{liapidevskii2018large} and \cite{gavrilov2013influence}. In this section we show evidence of their existence in the context of the Euler equations.

 One multi-hump ESW branch is shown in figure \ref{fig:H2_015_branch} with $H_2=0.15$ and $\Delta_1=\Delta_2=0.01$ for both the full Euler and $\MCC$ equations, where the parameter $Q$ on the $y$-axis measures the volume of the perturbation, given by
\begin{equation}
Q=\int_{-\infty}^0 \lvert \zeta_1 \rvert + \lvert\zeta_2\rvert \, \mathrm{d}x    .
\end{equation}
We use $Q$ as a bifurcation parameter rather than $A$ since the maximum displacement of a multi-hump solution is not necessarily at $x=0$. 
\begin{figure}
	\centering
	\begin{overpic}[scale=1]{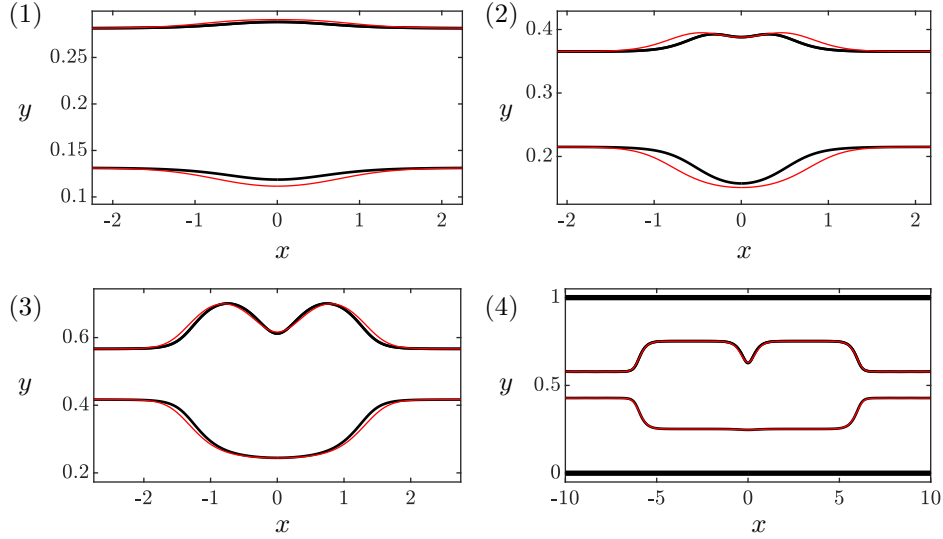}
		\put(0,56){$(1)$}
		\put(50,56){$(2)$}
		\put(0,25){$(3)$}
		\put(50,25){$(4)$}
		\put(1,46){$y$}
		\put(51.5,46){$y$}
		\put(1,17){$y$}
		\put(51.5,17){$y$}
		\put(28,31){$x$}
		\put(76.5,31){$x$}
		\put(28,2){$x$}
		\put(78,2){$x$}
	\end{overpic}
	\caption{Solutions (1)--(4) from figure \ref{fig:H2_015_branch}. The red curves are the $\MCC$ solutions, while the black curves are the corresponding Euler solution with the same value of $H_3$. The speeds of the solutions (1)--(4) are $c=0.02396$, $c=0.02772$, $c=0.03498$, and $c=0.03536$ for the $\MCC$ system and $c=0.02374$, $c=0.02764$, $c=0.03507$, and $c=0.03536$ for the Euler equations. 
	\label{fig:H2_015_soln}}
\end{figure}
\begin{figure}
	\centering
	\begin{overpic}[scale=1]{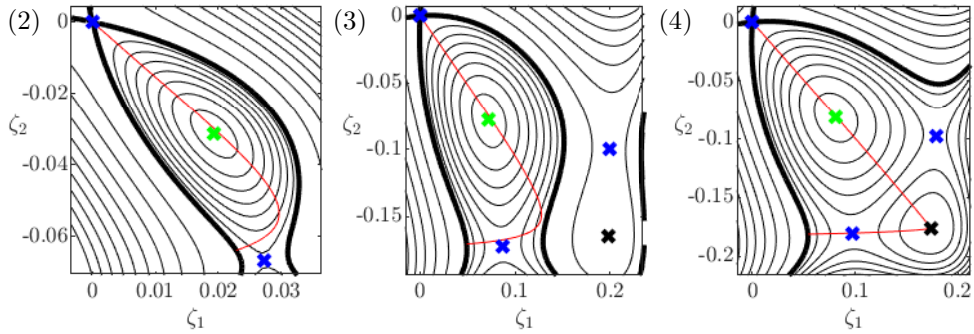}
		\put(0,31){$(2)$}
		\put(33,31){$(3)$}
		\put(66,31){$(4)$}
	\end{overpic}
	\caption{$\MCC$ solutions (2)--(4) from figure \ref{fig:H2_015_soln}, projected onto the $(\zeta_1,\zeta_2)$-plane. Black curves are level sets of $V$, where the bold black curve is $V=0$. The red curve is the trajectory of the solution, originating at the origin and bounded by $V=0$. The green, blue, and black crosses are minimum, saddle, and maximum critical points of $V$ respectively.  \label{fig:H2_015_projection}}
\end{figure}
\begin{figure}
	\centering
	\begin{overpic}[scale=1]{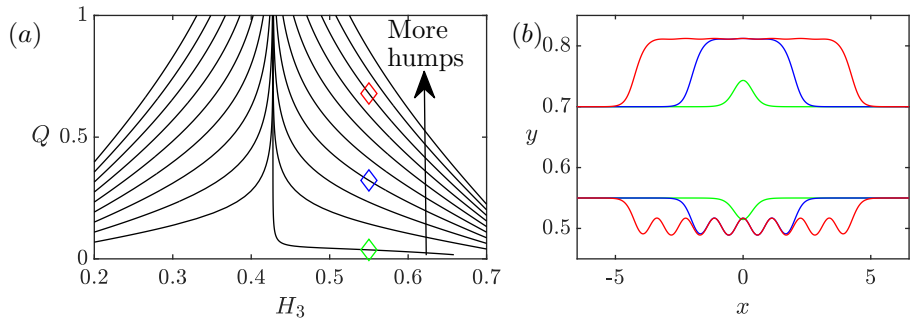}
	\put(0,30){$(a)$}
	\put(53,30){$(b)$}
	\put(40,30.5){More}
	\put(40,27.5){humps}
	\end{overpic}
	\caption{Panel $(a)$ shows branches of $\MCC$ ESWs with $H_2=0.15$ and $\Delta_1=\Delta_2=0.01$. The green, blue, and red diamonds lie on $H_3=0.55$, and these solutions are shown in panel $(b)$. The green solution has a speed $c= 0.02916$, while the blue and red solutions have almost the same speed $c=0.03126$. Solutions on branches with larger $Q$ are broader, with additional oscillations on the broadened section. \label{fig:newfig30}}
\end{figure}
The solutions (1)--(4) from figure \ref{fig:H2_015_branch} are shown in physical space in figure \ref{fig:H2_015_soln}. It can be seen that at with increasing values of $Q$ the waves broaden. However, unlike the single-humped ESWs seen in section \ref{section:results_symbrk}, the broadened section is not flat. To better understand this, we consider the projection of the $\MCC$ solutions onto the $(\zeta_1,\zeta_2)$-plane, as shown in figure \ref{fig:H2_015_projection}. As the amplitude of the solution increases, the trajectory gets closer to the maximum critical point, shown by the black cross. The single-humped ESW branch with the same values of $H_2$ and $\Delta_i$ has a limiting solution which is a heteroclinic orbit from the origin to the maximum critical point. Remarkably, the multi-hump branch from figure \ref{fig:H2_015_branch} appears to approach a solution (see solution (4)) which consists of a quasi-heteroclinic orbit to the maximum critical point combined with a quasi-homoclinic orbit originating at the maximum critical point: in other words a solitary wave riding on a conjugate state.

One can smoothly, via continuation, get from an $\MCC$ solution along this branch to a $(1,2)$ compacton by decreasing the value of $H_2$. One can also reduce the value of $H_2$ for the Euler solutions, but we found computing solutions with $H_2<0.05$ difficult due to stiffness in the numerical method.   

More multi-hump ESW branches with $H_2=0.15$ and $\Delta_1=\Delta_2$ can be found, such as the $\MCC$ branches shown in figure \ref{fig:newfig21}. All the branches shown have a vertical asymptote at $H_3\approx 0.4276$. As indicated by the arrow in the figure, solutions on branches with larger $Q$ are characterised by a broadened wave with additional oscillations on the broadened profile. For example, fixing $H_3=0.55$, we plot the solutions indicated with the diamonds in panel $(b)$.  The green single-humped profile is akin to those seen in section \ref{section:results_symbrk}. Meanwhile, the blue and red solutions are multi-humped ESWs. The oscillations on the broadened section are seen clearly on the lower interface, while the oscillations on the upper interface are less visible. Only when $H_2=0$ do the oscillations on the top interface disappear, resulting in an $(n,1)$ compacton.  As $H_2\rightarrow 0$, the blue and red solutions approach (4,1) and (8,1) compactons. Our numerical results suggest there are solutions with an arbitrary number of such oscillations, and can be found by continuation of a compacton solution. We find this same behaviour for the Euler system, although the limiting configuration cannot be reached for solutions with many humps due to the large computational domains required.

\begin{figure}
	\centering
	\begin{overpic}[scale=1]{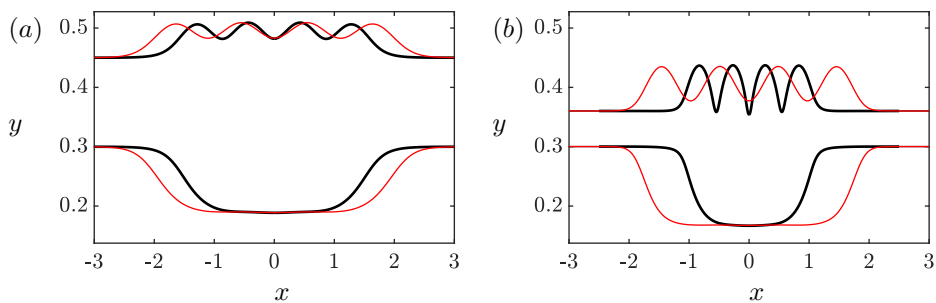}
		\put(-1,27){$(a)$}
		\put(50,27){$(b)$}
		\put(-1,17){$y$}
		\put(50,17){$y$}
		\put(27,-1){$x$}
		\put(77,-1){$x$}
	\end{overpic}
	\caption{ESWs for the $\MCC$ (red) and Euler (black) equations with $H_3=0.3$ and $\Delta_1=\Delta_2=0.01$. The solutions in panel $(a)$ have $H_2=0.15$, where the $\MCC$ and Euler speeds are $c=0.03112$ and $c=0.03109$ respectively. Meanwhile, panel $(b)$ has $H_2=0.06$ with $\MCC$ and Euler speeds of $c=0.02895$ and $c=0.02888$ respectively         \label{fig:14compare}}
\end{figure}
\begin{figure}
    \centering
    \begin{overpic}[scale=1]{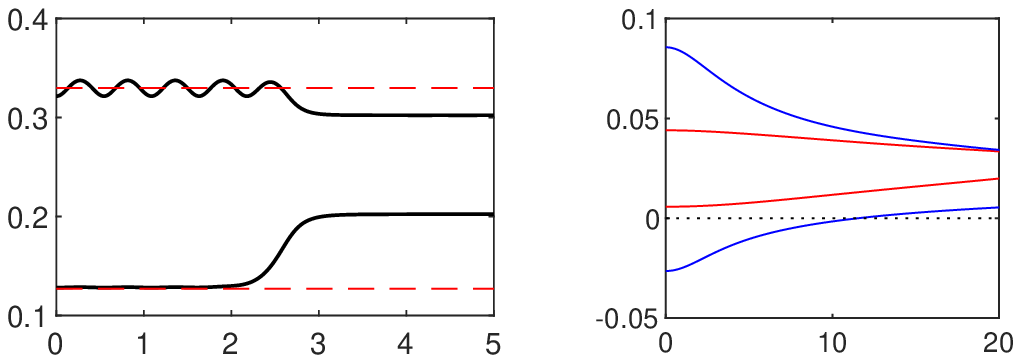}
    \put(0,21){$y$}
     \put(30,0){$x$}
      \put(56,21){$U$}
     \put(82,0){$k$}
     \put(0,34){$(a)$}
     \put(56,34){$(b)$}
    \end{overpic}
    \caption{Panel $(a)$ shows an Euler ESW with $H_2=0.1$, $H_3=0.202306$, $\Delta_1=\Delta_2=0.01$, and $c=0.0256299$. The red curves are a saddle conjugate state of the far-field, with values $\hat{\zeta_2}=-0.0752539$, $\hat{\zeta_1}=0.0276113$, and $c=0.0256230$. Panel $(b)$ shows the linear dispersion relation of a wave travelling to the right with speed $U$ on the conjugate state. Linear modes in which the interface displacements are the same sign are in blue, while those with opposite signed interface displacements are in red. It can be seen there is a standing linear wave $(U=0)$ with wavenumber $k=11.54$, in excellent agreement with the humps on the ESW with $k=11.53$. } 
    \label{fig:N1}
\end{figure}
As with the single-humped ESWs, we found that the $\MCC$ system becomes a poorer approximation for the Euler equations as $H_2$ decreases. For example, in figure \ref{fig:14compare}, we compare $\MCC$ and Euler multi-hump profiles for $H_2=0.15$ and $H_2=0.06$ in panel $(a)$ and $(b)$ respectively. It can be seen that the wavelength and amplitude of oscillations at the center of the wave have worse agreement for the solutions in panel $(b)$.

We now relate these solutions back to the discussion from section 3. The multi-hump solutions observed here have the same characteristics as the solution seen in figure \ref{fig:saddlesaddlenew}. The origin is a saddle of $V$, and hence oscillations in the tail will be observed except for special values of the parameters. These solutions are a homoclinic orbit starting at the origin which approach a periodic mode-1 wave. This mode-1 wave, for the MCC3 system, is in the vicinity of a saddle critical point. These solutions have speeds slightly larger than that of a saddle mode-2 conjugate state. Noting that saddle mode-2 conjugate states exist for all parameters except at criticality (e.g. figure \ref{fig:CS_op}), multi-hump solutions appear to be extremely common. The limiting single humped solutions seen in section \ref{section:results_symbrk}, such as figure \ref{fig:newfig16}$(b)$, are akin to the solution in figure \ref{fig:saddlemaxnew}. The origin is still a saddle, and hence oscillations in the tail must be avoided by careful choice of parameters, but now the conjugate state is a maximum, and so limiting solutions are wavefronts.

\begin{figure}
   \centering
   \begin{overpic}[scale=1]{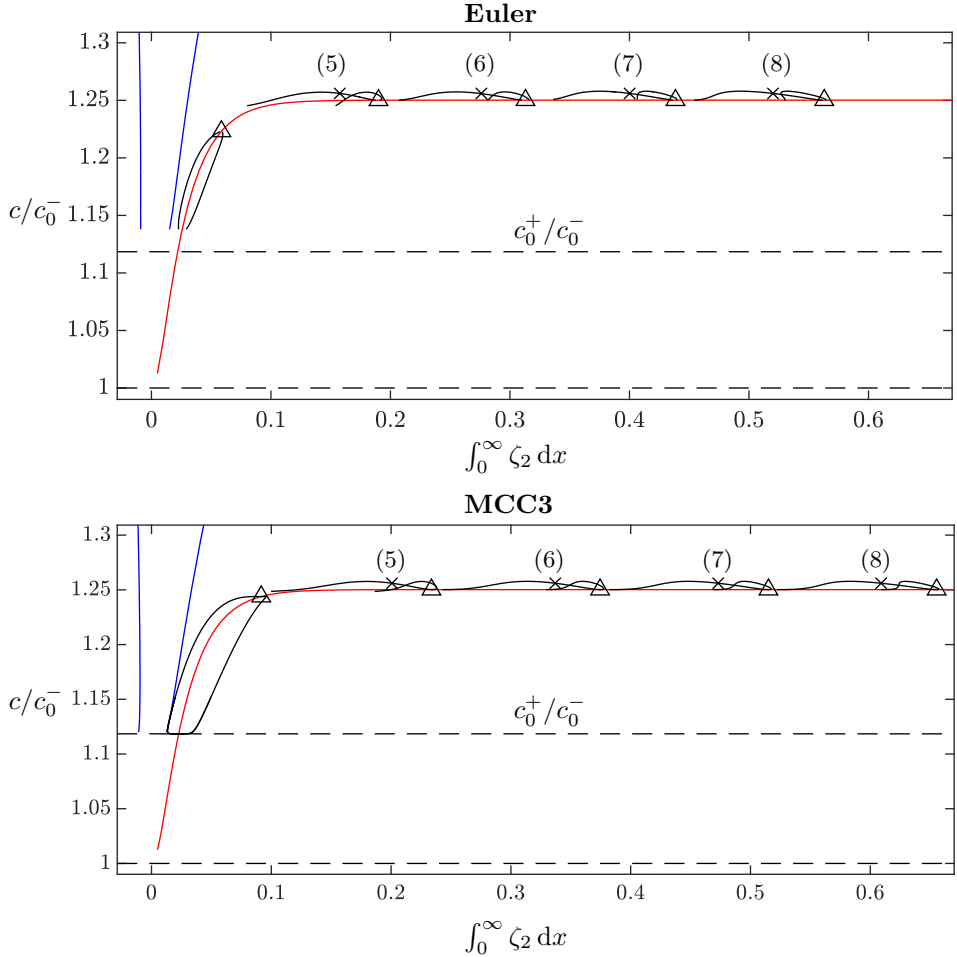}
   \put(45,50){$\int_0^{\infty} \zeta_2 \, \mathrm{d} x$}
   \put(45,1){$\int_0^{\infty} \zeta_2 \, \mathrm{d} x$}
   \put(-1,25){$c/c_0^-$}
   \put(-1,75){$c/c_0^-$}
   \put(45,45){\textbf{MCC3}}
   \put(45,95){\textbf{Euler}}
   \put(30,90){\small (5)}
   \put(45,90){\small (6)}
   \put(60,90){\small (7)}
   \put(75,90){\small (8)}
   
   \put(36,39.5){\small (5)}
   \put(52,39.5){\small (6)}
   \put(69,39.5){\small (7)}
   \put(85,39.5){\small (8)}
   
   \put(50,73){$c_0^+/c_0^-$}
   \put(50,24){$c_0^+/c_0^-$}
      
   \end{overpic}
    \caption{Boussinesq solution branches with $H_2=0.8$, $H_1=H_3=0.1$, and $\Delta_1=\Delta_2=0.01$. The top panel is for the Euler system, while the bottom is for $\MCC$. The two blue curves are mode-1 solitary wave branches. The red curve is the symmetric Boussinesq mode-2 branch, while the black curves are mode-2 branches which bifurcate from the symmetry branch. The triangles are bifurcation points. Only the first five bifurcating branches are shown. The solutions $(5)$--$(8)$ corresponding to the crosses are plotted in figures \ref{fig:con_soln}. 
    \label{fig:concave_branches}}
\end{figure}

Information regarding critical points of a potential only applies to the MCC3 equations, yet the solution behaviour is remarkably similar to the Euler system. Therefore, we provide another interpretation of large amplitude multi-hump waves. The humps are can be seen as mode-1 periodic standing waves which lie on a solution branch which bifurcates with zero amplitude from the conjugate state! The wave properties of these oscillations can be predicted by linear theory when they are of small amplitude.
This is demonstrated in figure \ref{fig:N1}, where panel $(a)$ shows an Euler multi-hump ESW (only the $x>0$ portion of the profile is shown). In red, we plot the mode-2 saddle conjugate state corresponding to the values of $H_i$ and $\rho_i$. The wavenumber of one of the mode-1 oscillations on the broadened section is $k=11.5288$. Taking values of $\hat{H_i}$ and $\hat{u_i}$ (see figure \ref{fig:CS}) for the conjugate state, one can linearise the system by seeking wave forms travelling at a constant speed $U$. The dispersion relation is omitted here, but can be found in \cite{baines1997topographic} and has four branches.
The leftward and rightward travelling waves are no longer equivalent due to the different values of $\hat{u_i}$ across each layer. It can be seen that a linear standing wave ($U=0$) exists for $k=11.5430$, in strong agreement with the mode-1 oscillations on the ESW. The discrepancy arises due to the nonlinearity of the mode-1 oscillations, and that the speed of the ESW is slightly larger than that of the conjugate state. Another way of stating the above is that the mode-1 periodic wave on the conjugate state travels at the same speed as the ESW.

\begin{figure}
    \centering
    \begin{overpic}[scale=1]{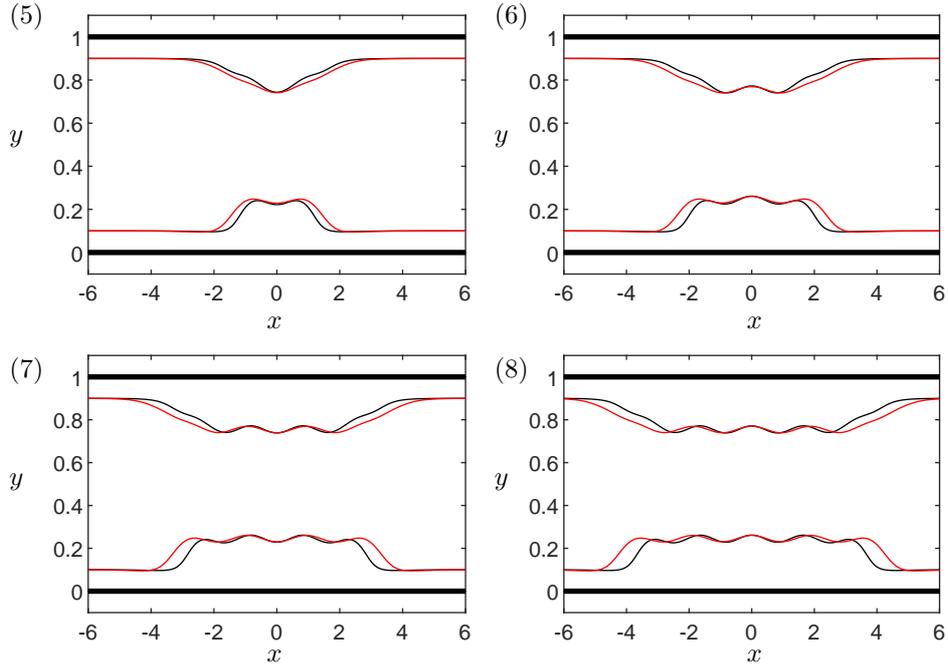} 
    \put(2,67){$(5)$}
    \put(51,67){$(6)$}
    \put(2,31){$(7)$}
    \put(51,31){$(8)$}
    
    \put(2,55){$y$}
    \put(51,55){$y$}
    \put(2,20){$y$}
    \put(51,20){$y$}
    
    \put(28,36){$x$}
    \put(76.5,36){$x$}
    \put(28,2){$x$}
    \put(76.5,2){$x$}
    
    \end{overpic}
    \caption{Solutions $(5)$--$(8)$ from figures \ref{fig:concave_branches}. The black curves are the Euler solution, while the red curves are $\MCC$. The solutions all have the same speed $c=0.03552$, which is greater than $c_0^-= 0.0316$.} 
    \label{fig:con_soln}
\end{figure}

\begin{figure}
    \centering
    \begin{overpic}[scale=1]{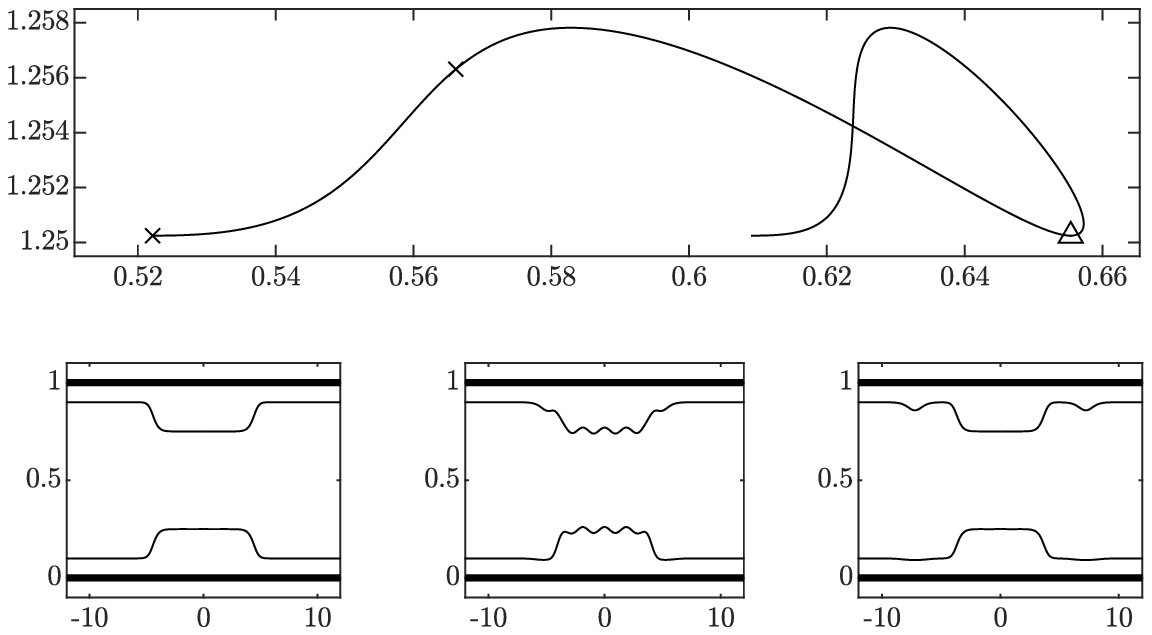} 
    \put(0,43){$\frac{c}{c_0^-}$}
    \put(52,28){$\int_0^{\infty}\zeta_2 \, \mathrm{d} x  $}
      \put(20.5,1){$x$}
      \put(53,1){$x$}
      \put(84.5,1){$x$}
      \put(2,15){$y$}
      \put(35,15){$y$}
      \put(67,15){$y$}
      
      \put(2,23){$(a)$}
      \put(34,23){$(b)$}
      \put(68,22){$(c)$}
      
      \put(13.5,37.5){$(c)$}
      \put(40.5,45){$(b)$}
      \put(88,37){$(a)$}
    \end{overpic}
    \caption{A single $\MCC$ branch from figure \ref{fig:concave_branches} showing the bifurcation point, and solutions along the branch.}
    \label{fig:con_singlebranch}
\end{figure}
\begin{figure}
    \centering
    \begin{overpic}[scale=1]{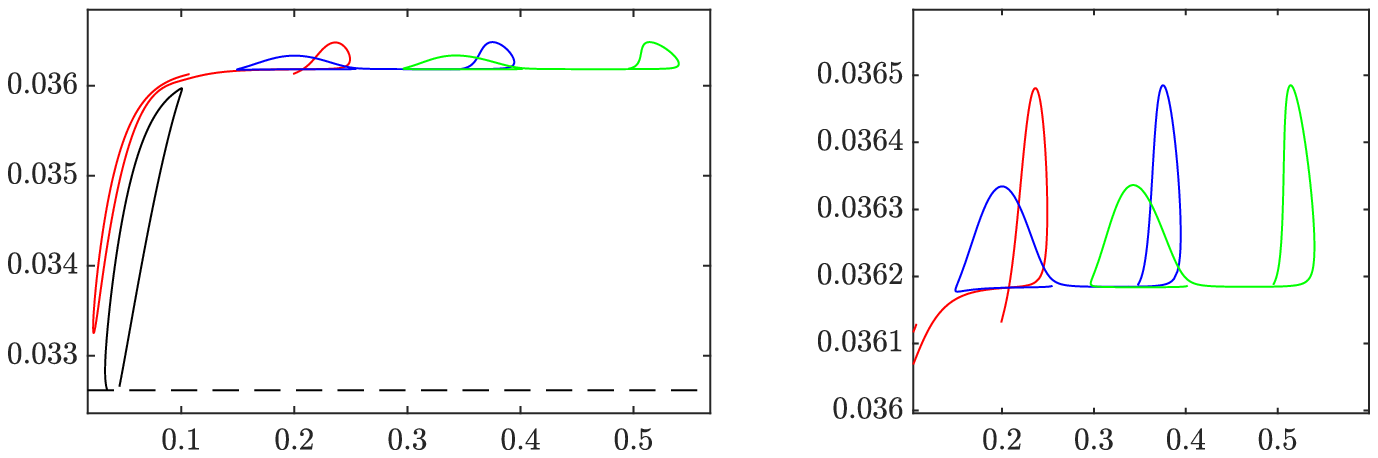} 
    \put(27,-1){$\int_0^{\infty} \zeta_2 \, \mathrm{d}x$}
    \put(77,-1){$\int_0^{\infty} \zeta_2 \, \mathrm{d}x$}
    \put(0,18){$c$}
    \put(55,18){$c$}
     \put(0,30){$(a)$}
    \put(55,30){$(b)$}
    \put(27,7){$c_0^+$}
    \end{overpic}
    \caption{Boussinesq  $\MCC$ mode-2 solitary wave solution branches with parameter values $\Delta_1=0.011$, $\Delta_2=0.01$, $H_2=0.8$, and $H_1=H_3=0.1$. Different colours are used to distinguish separate branches. Panel $(b)$ shows a blow-up of the red, blue, and green branch from panel $(a)$. }
    \label{fig:con_symbreak}
\end{figure}

\subsection{Concave waves outside the linear spectrum}\label{section:concave}
All of the mode-2 solitary waves considered so far have speeds less than $c_0^+$, and are hence ESWs. However, it is not always the case that mode-2 waves must be within the linear spectrum. This is predicted by the fact that mode-2 conjugate states may have speeds exceeding $c_0^+$, provided that the value of $H_2$ is sufficiently large (see figure \ref{fig:newfig_CS_bous}(c)). 
The mode-2 conjugate state with $c>c_0^+$ is a saddle of the $\MCC$ potential. We have seen already in the previous section how large amplitude ESWs can have a broadened section with periodic-like orbits in the vicinity of a  non-trivial saddle point, resulting in multi-hump solitary waves. Such solutions are also found outside the linear spectrum, as discussed below. For simplicity, we will restrict our attention to the Boussinesq system.

We begin by exploring solutions with $H_2=0.8$, $H_1=H_3$, and $\Delta_1=\Delta_2$. 
A bifurcation diagram is shown in figures \ref{fig:concave_branches} for the Euler and $\MCC$ systems in the upper and lower panels. The horizontal axis is the volume of the perturbation of the lower interface over half the domain. The agreement, like for convex waves, is qualitatively good but can be quantitatively poor. 
The two blue curves correspond to two mode-1 solitary wave branches which bifurcate from $c=c_0^+$. Since this is the symmetric Boussinesq configuration, one branch can be recovered from the other by the mapping $g\rightarrow -g$, $\zeta_1\rightarrow -\zeta_2$, $\zeta_2\rightarrow -\zeta_1$. The red branch is mode-2 solitary waves with $\zeta_1=-\zeta_2$, which bifurcate from $c=c_0^-$ and approach a tabletop solitary wave. As expected, the limiting solution is a heteroclinic orbit from the origin to the mode-2 conjugate state of the system.
 Along the branch there are bifurcation points represented by triangles. The $\zeta_1=-\zeta_2$ symmetry is broken along the bifurcating branches, which are shown by the black curves. In fact, two new branches bifurcate from each bifurcation point, but they are related to each other by the aforementioned mapping for the two mode-1 branches.


Solutions along the bifurcating branches are multi-humped solitary waves. Figure \ref{fig:con_soln} shows the solutions (5)--(8) from figures \ref{fig:newfig30}, where the Euler and $\MCC$ profiles are shown in black and red respectively. For comparison, the speeds are kept the same across all the solutions. It can be seen that the interfaces have additional oscillations as one considers the sequence of bifurcating branches. Our numerical exploration suggests infinitely many of these bifurcating branches exist.
We plot a single bifurcating branch for the $\MCC$ system in figure \ref{fig:con_singlebranch}. At the bifurcation point, the solution is a tabletop solitary wave, as shown by panel $(a)$. As the speed of the wave increases, oscillations appear on both interfaces, as seen by solution $(b)$. Finally, at the end of the branch, the solution is again a tabletop solitary wave, but with a mode-1 solitary wave either side of the broad mode-2 solution. The mode-1 solitary wave is the solution from the mode-1 branch with the same speed. Moving along the branch beyond the solution $(c)$, one finds the mode-1 waves can be made to be arbitrarily far from the mode-2 wave, but the speed of the wave (and the profiles of the mode-1 and mode-2 waves) remain unchanged.

Next, we consider a solution branch with non-symmetric stratification, by setting $H_2=0.8$, $H_1=H_3$, $\Delta_1=0.01$ and $\Delta_2=0.011$. For simplicity, we shall consider the $\MCC$ system only. The solution space for mode-2 solitary waves is shown in figure \ref{fig:con_symbreak}, where we plot each branch in a different colour. There are two key differences with the solution space for the symmetric Boussinesq configuration. First, there is no longer a branch satisfying $\zeta_1=-\zeta_2$, and hence none of the branches with $c>c_0^+$ smoothly enter the linear spectrum while remaining truly localised. In the linear spectrum, one may still find ESWs, but they are isolated points along branches of GSW, as seen for convex waves in section \ref{section:results_symbrk}. The second difference is that there is no sequence of bifurcation points. Instead, the branches of solutions are isolated: where they cross no longer corresponds to a bifurcation point. This being said, numerics still suggest there is an infinite sequence of multi-humped solution branches with an increasing number of oscillations on the broadened section.

 \begin{figure}
     \centering
     \begin{overpic}[scale=1]{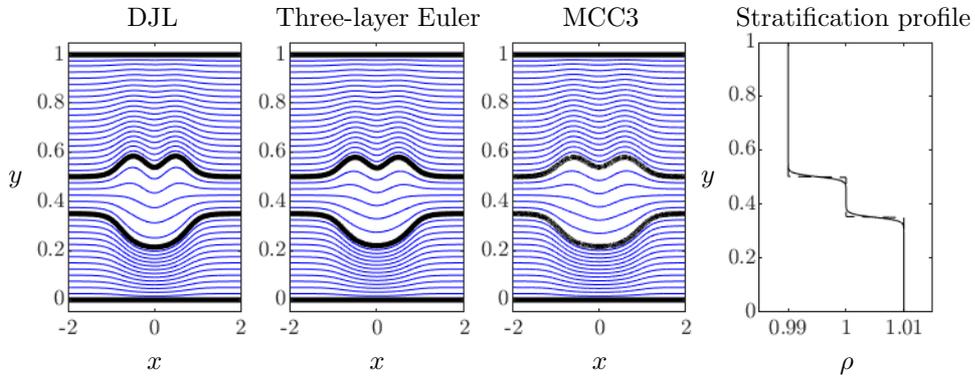}
     \put(18,0){$x$}
     \put(40.5,0){$x$}
     \put(63,0){$x$}
     \put(88,0){$\rho$}
     \put(74,19){$y$}
     \put(4,19){$y$}
     \put(16,35){DJL}
     \put(31,35){Three-layer Euler}
     \put(60,35){MCC3}
     \put(77.5,35){Stratification profile}
     \end{overpic}
   \caption{Mode-2 ESW in a continuously stratified fluid with two pycnoclines. Boussinesq multi-hump solutions with internal streamlines are shown for the continuously stratified DJL system and the layered $\MCC$ and Euler systems. Internal streamlines are plotted in blue, while the streamlines corresponding to the interface of the three-layer model are in black. The continuous density stratification used in the DJL equation is given by  $\rho(y) = \rho_3 + \Delta_2 \lrr{1+\tanh\lrr{(y-H_3)/d}}/2 +\Delta_1 \lrr{1+\tanh\lrr{(y-H_3-H_2)/d}}/2$ with $d=80$ and $\Delta_1=\Delta_2=0.01$ and is shown by the solid curve in the right-most panel, while the dashed curve shows the discontinuous density profile of the three-layer models (with the same values of $\Delta_i$).  All solutions are for $H_3=0.35$, $H_2=0.15$, and $H_1=0.5$. The speeds of the DJL, three-layer Euler, and MCC3 solutions  are  $c=0.03191$, $c=0.03298$, $c=0.03301$, respectively. It was found that increasing the value of $d$ has little effect on the streamlines for the DJL solution, but the speed of the wave increases, improving agreement with the discontinuously stratified three-layer solution. \label{fig:DJL}}
     \label{fig:my_label}
 \end{figure}

\section{Concluding remarks}
In this paper, we present a variety of steady three-layer mode-2 internal solitary waves. Using the criterion proposed by \cite{champneys2001embedded}, we establish numerically that truly localised mode-2 solitary waves can be embedded in the linear spectrum.
In addition to single-humped profiles, we find branches of exotic multi-hump solutions for both the fully nonlinear Euler equations and the $\MCC$ equations.
 For MCC3, this extends the work of \cite{liapidevskii2018large} and \cite{barros2020strongly}, who previously computed multi-humped waves within the strongly nonlinear regime, but limited to small depths of the intermediate layer. In fact, such solutions are revealed to be prolific throughout the parameter space.

The $\MCC$, being a long-wave strongly nonlinear model, is found to be in good agreement with the Euler system as long as the shallow water assumption is not invalidated. We have found that, even in cases where the the long wave approximation is not applicable, qualitative agreement still holds for large amplitude solutions, but significant quantitative differences can be found. It should be noted that the MCC3 travelling wave solutions are governed by a coupled system of second order ordinary differential equations, whose solutions are relatively simple to compute.

All mode-2 solutions found in this paper tend to broaden with increasing values of amplitude. 
In some cases, the broadened section is flat (i.e. tabletop solitary waves), while in others, numerous oscillations occur on the broadened section (i.e. multi-hump solitary waves).
We have seen that this relates entirely to the nature of the conjugate states of the system and their nature as critical points of MCC3. It is fascinating to note that the while this critical point analysis follows from the MCC3 system, the consequence it has on the solution space appears to hold for the full Euler system.
Our work also suggests that, except at mode-2 criticality, multi-hump waves occur across parameter space. Furthermore, mode-2 solitary waves were found with speeds greater than the linear long-wave speed of mode-1, predicted by the existence of a mode-2 conjugate state outside the linear spectrum.

The work in this paper is restricted to a three-layer model, which idealises a continuously stratified fluid with two sharp pyncoclines. The existence of ESWs in a continuous stratification is the subject of an ongoing investigation. We present here preliminary results, which demonstrate that the localised coherent structures found for the layered model in this paper still exist in the more realistic context of continuous stratification. As an example, we take a two-pycnocline $\tanh$ stratification profile given in \cite{lamb2000conjugate}
to find ESW solutions to the Euler equations in continuous stratification that, in steady form, are known as the  Dubriel-Jacotin-Long (DJL) equation \citep{DJ,L}. This is shown in figure \ref{fig:DJL}, where we plot a multihump ESW with comparable parameters for the $\MCC$, three-layer Euler and DJL system. Streamlines are included to demonstrate the similar vertical structure, where the streamlines for the $\MCC$ are recovered using the leading order solution of the model. The DJL equation is solved using a finite difference scheme with second-order central-difference approximations of derivatives.

The most pertinent question about these solutions regards their stability and generation. While layered models will suffer from Kelvin-Helmholtz instability due to a jump in densities, it may be possible that ESWs in continuously stratified fluids are stable. Indeed, a multi-hump mode-2 solitary wave was found experimentally by \cite{liapidevskii2018large} with two humps on the lower interface and one on the upper interface. The wave was generated with a lock-release mechanism, and propagated keeping its permanent form. They noted that non-symmetric mode-2 solitary waves waves would radiate energy via a mode-1 tail except `for special choice of stratification', in-line with the results of this paper which found that localised mode-2 solitary waves have delicate dependence on parameters. The solutions may also be attractors, arising after wave collision or wave interaction with topography. These are topics of future research.

We have demonstrated the rich bifurcation structure of higher-mode solitary waves and presented the first example of non-trivial ESWs within fully nonlinear theory. Our numerical solutions should also spawn interesting analytic conjectures regarding their existence. This paper has focused primarily on solutions in which the stratification is  weak  ($\Delta_1\approx \Delta_2 \ll 1$). Owing to the large number of parameters in the problem, a complete description of the solution space for steady waves is still to be found.
 Previous computations by \cite{rusaas2002solitary} for stronger stratification have found existence of overhanging solutions for three-layer mode-1 waves. Whether such behaviour could be found for mode-2 waves is unknown.

 \vspace{4mm}
\noindent \textbf{Declaration of interests.} The authors report no conflict of interest.
 
\bibliography{three_layer_bib}  
\bibliographystyle{abbrvnat}

\begin{appendix}

\section{Numerical method for Euler equations}\label{section:num_meth}
To solve the system of equations described in section \ref{section:form}, we reformulate the problem such that $x$ and $y$ are functions of the independent variables $\phi_3$ and $\psi_3$. The desire to solve this system of equations in the potential space is motivated by the fact that the unknown interfaces are isolines of the streamfunctions  $\psi_i$. The unknowns are expressed in terms of integrals concerning boundary values via Cauchy's integral formula. This method has been adopted by many authors for the two-layer model (for example, \cite{turner1986limiting}
). We note that this method is very similar to that used by \cite{rusaas2002solitary} for the three-layer model. However, they remain in the physical space and resolve boundary integrals seeking the unknown velocities in terms of $x$ and $y$. The advantage of working in the potential space is the reduction in the number of unknowns, as demonstrated below.

Consider first the bottom fluid layer. Denoting $\phi=\phi_3$, we can write the complex potential $f_3$ in the lower fluid layer as $f_3=\phi+i\psi_3$. Without loss of generality, choose $\psi_3=0$ on the bottom wall, and $\phi=0$ at $x=0$. The potential space of the bottom fluid layer, and it's reflection across the wall $\psi_3=0$, is given by
\begin{equation}
\Omega^f_3 = \{(\phi,\psi_3): \phi\in[-\tilde{c}\lambda/2,\tilde{c}\lambda/2], \psi_3\in(-2Q_3,0)\},
\end{equation}
where we denote the flux of fluid in the bottom fluid layer as $Q_3$.
We conformally map $\Omega^f_3$ to an annulus in the $t$-plane via the mapping
\begin{equation}\label{eq:t3}
t = \exp\left(-\frac{kif_3}{\tilde{c}} \right).
\end{equation}
We parameterise the perturbation to the uniform stream of the lower interface as $(X(\phi),Y(\phi))$ and the upper interface as $(\tilde{X}(\phi),\tilde{Y}(\phi))$, where
\begin{equation}
\int\displaylimits_0^{c\lambda/2} Y(\phi) \, \mathrm{d} \phi = 0, \hspace{1cm} \int\displaylimits_0^{c\lambda/2} \tilde{Y}(\phi) \, \mathrm{d} \phi  = 0. 
\end{equation}
The function $F=x_\phi + 1/\tilde{c} + iy_\phi$, where subscripts denote partial differentiation, is an analytic function of $f_3$, and hence an analytic function of $t$. Applying Cauchy's integral formula to the function $F$ over the annulus in the $t$-plane, where we denote the boundary of the annulus traversed counterclockwise as $C$, one finds that 
\begin{equation}\label{eq:CIF}
F(t_0) = \frac{1}{\pi i} \oint_C \frac{F(t)}{t-t_0} \, \mathrm{d} t,
\end{equation}
where $t_0$ is taken on the boundary of the annulus.
Making use of (\ref{eq:t3}), and taking real parts, the above integral gives
\begin{align}\label{eq:int1}
\frac{\tilde{c}\lambda}{2}\left(X_\phi(\phi_0) + \frac{1}{\tilde{c}}\right)&=  \frac{1}{2}
\int\displaylimits_{-\tilde{c}\lambda/2}^{\tilde{c}\lambda/2} Y_\phi \cot\left(\frac{\Theta_-}{2}\right) \, \mathrm{d} \phi \nonumber \\
&- \int\displaylimits_{-\tilde{c}\lambda/2}^{\tilde{c}\lambda/2} \frac{\left(X_\phi + \frac{1}{\tilde{c}}\right)\left(1-R_3 \cos\Theta_-\right)}{1+R_3^2 - 2R_3 \cos\Theta_-}  \, \mathrm{d} \phi \nonumber \\
&+ \int\displaylimits_{-\tilde{c}\lambda/2}^{\tilde{c}\lambda/2} \frac{Y_\phi \sin\Theta_-}{1+R_3^2 - 2R_3 \cos\Theta_-}  \, \mathrm{d} \phi .
\end{align} 
Here, $\Theta_\pm = k(\phi_0\pm\phi)/c$ and $R_3=e^{2Q_3 k/\tilde{c}}$. The above integral expresses $X_\phi(\phi_0)$ at some point $\phi_0$ on the interface in terms of integrals concerning boundary values of $X_\phi$ and $Y_\phi$.

Next, we consider the intermediate fluid layer. We write $\phi_2=h(\phi)$, and denote the flux in the layer as $Q_2$. Without loss of generality, let $\psi_2=0$ on the lower interface $y=\eta_2$. Hence, the potential space of the intermediate layer $f_2=h(\phi)+\psi_2$ is given by
\begin{equation}
\Omega^f_2 = \{(h(\phi),\psi_2): \phi\in[-\tilde{c}\lambda/2,\tilde{c}\lambda/2], \psi_2\in(0,Q_2)\}.
\end{equation}
The far-field condition \eqref{eq:farfield1} requires that $h$ satisfies
\begin{equation}\label{eq:hfarfield}
h_\phi\left(\frac{\pm\tilde{c}\lambda}{2}\right) = 1.
\end{equation}
One can map the above space to an annulus in the $s$-plane via the mapping
\begin{equation}
s = \exp\left(\frac{kif_2}{\tilde{c} h_0} \right).
\end{equation}
where 
\begin{equation}\label{eq:h0}
h_0 = \frac{h(\tilde{c}\lambda/2) - h(0) }{\tilde{c}\lambda/2}.
\end{equation}
The parameter $h_0$ is a measure of the wavelength averaged shear between the bottom and intermediate layer. When $h_0=1$, then $\phi_3$ and $\phi_2$ vary the same over one wavelength, and hence the average shear is zero. 
Typically, $h_0$ is different from unity. Again solving Cauchy's integral formula on the region bounded by the annulus in the $s$-plane, one can derive two boundary integral equations. First we find that
\begin{align}\label{eq:int2}
\frac{\tilde{c}h_0\lambda}{2}\left(\frac{X_\phi(\phi_0)}{h_\phi(\phi_0)} + \frac{1}{\tilde{c}h_0}\right)&=  - \frac{1}{2}
\int\displaylimits_{-\tilde{c}\lambda/2}^{\tilde{c}\lambda/2} Y_\phi \cot\left(\frac{\alpha_-}{2}\right) \, \mathrm{d} \phi \nonumber \\
&- \int\displaylimits_{-\tilde{c}\lambda/2}^{\tilde{c}\lambda/2} \frac{\left(\tilde{X}_\phi + \frac{h_\phi}{\tilde{c}h_0}\right)\left(1-R_2 \cos\alpha_-\right)}{1+R_2^2 - 2R_2 \cos\alpha_-}  \, \mathrm{d} \phi \nonumber \\
&+ \int\displaylimits_{-\tilde{c}\lambda/2}^{\tilde{c}\lambda/2} \frac{\tilde{Y}_\phi \sin\alpha_-}{1+R_2^2 - 2R_2 \cos\alpha_-}  \, \mathrm{d} \phi .
\end{align} 
Here, $R_2=e^{Q_2 k/\tilde{c}}$ and $\alpha_\pm=k(h(\phi_0)\pm h(\phi))/(\tilde{c}h_0)$. Second, we get 
\begin{align}\label{eq:int3}
\frac{\tilde{c}h_0\lambda}{2}\left(\frac{\tilde{X}_\phi(\phi_0)}{h_\phi(\phi_0)} + \frac{1}{\tilde{c}h_0}\right)&=  - \frac{1}{2}
\int\displaylimits_{-\tilde{c}\lambda/2}^{\tilde{c}\lambda/2} \tilde{Y}_\phi \cot\left(\frac{\alpha_-}{2}\right) \, \mathrm{d} \phi \nonumber \\
&- \int\displaylimits_{-\tilde{c}\lambda/2}^{\tilde{c}\lambda/2} \frac{\left(X_\phi + \frac{h_\phi}{\tilde{c}h_0}\right)\left(1-R_2 \cos\alpha_-\right)}{1+R_2^2 - 2R_2 \cos\alpha_-}  \, \mathrm{d} \phi \nonumber \\
&- \int\displaylimits_{-\tilde{c}\lambda/2}^{\tilde{c}\lambda/2} \frac{Y_\phi \sin\alpha_-}{1+R_2^2 - 2R_2 \cos\alpha_-}  \, \mathrm{d} \phi .
\end{align}
Finally, denoting $\phi_1=g(\phi)$ and the flux in the upper layer as $Q_1$, we map the upper interface and it's reflection across the top boundary in the $(g,\psi_1)$-space to an annulus. Following the same methodology as above, and upon assuming $\psi_3=0$ on the upper interface $y=\eta_1$, we get
\begin{align}\label{eq:int4}
\frac{\tilde{c}g_0\lambda}{2}\left(\frac{\tilde{X}_\phi(\phi_0)}{g_\phi(\phi_0)} + \frac{1}{\tilde{c}g_0}\right)&=  - \frac{1}{2}
\int\displaylimits_{-\tilde{c}\lambda/2}^{\tilde{c}\lambda/2} \tilde{Y}_\phi \cot\left(\frac{\beta_-}{2}\right) \, \mathrm{d} \phi \nonumber \\
&- \int\displaylimits_{-\tilde{c}\lambda/2}^{\tilde{c}\lambda/2} \frac{\left(\tilde{X}_\phi + \frac{g_\phi}{\tilde{c}g_0}\right)\left(1-R_1 \cos\beta_-\right)}{1+R_1^2 - 2R_1 \cos\beta_-}  \, \mathrm{d} \phi \nonumber \\
&- \int\displaylimits_{-\tilde{c}\lambda/2}^{\tilde{c}\lambda/2} \frac{\tilde{Y}_\phi \sin\beta_-}{1+R_1^2 - 2R_1 \cos\beta_-}  \, \mathrm{d} \phi .
\end{align}  
Here, $R_1= \exp(2kQ_1/(\tilde{c}g_0))$, $\beta_\pm=k(g(\phi_0)\pm g(\phi))/(\tilde{c}g_0)$, and
\begin{equation}\label{eq:g0}
g_0 =  \frac{g(\tilde{c}\lambda/2) - g(0) }{\tilde{c}\lambda/2}.
\end{equation}
The far-field condition \eqref{eq:farfield2} requires that
\begin{equation}\label{eq:gfarfield}
g_\phi\left(\frac{\pm\tilde{c}\lambda}{2}\right) = 1.
\end{equation}
It is left to satisfy the continuity of pressure condition (\ref{eq:Bern}). We can rewrite these equations as
\begin{align}
h_\phi(\phi) &= \left[ \rho_3 + \frac{2}{ q^2} \left( \Delta_2 Y - B_2 \right)                  \right]^{1/2}, \label{eq:h}\\
g_\phi(\phi) &= \left[ \frac{1}{\rho_1} h_\phi^2   +  \frac{2}{\rho_1\tilde{q}^2} \left(\Delta_1\tilde{Y} - B_1 \right)           \right]^{1/2}, \label{eq:g}
\end{align}
where 
\begin{equation}
q = (X_\phi^2 + Y_\phi^2)^{-1/2}, \hspace{1cm} \tilde{q} = (\tilde{X}_\phi^2 + \tilde{Y}_\phi^2)^{-1/2}.
\end{equation}
Similar expressions are found when using the Boussinesq approximation.

We wish to solve the reformulated problem numerically. We use the assumed symmetry about $x=0$ such that we only have to solve the problem over half a wavelength. We discretise $\phi\in[-\tilde{c}\lambda/2,0]$ into $N+1$ mesh points $\phi_I$ as follows
\begin{equation}
\phi_I = -\frac{\tilde{c}\lambda}{2} \left[\frac{N+1-I}{N}\right], \hspace{1cm} I=1,\cdots,N+1.
\end{equation}
The midpoints are give by $\phi_I^M = (\phi_{I} + \phi_{I+1})/2$. We express some of the unknowns as a Fourier series in $\phi$,
\begin{align}
Y(\phi)  &= \sum_{n=1}^{N-1} a_n \cos\left(\frac{kn}{\tilde{c}}\phi \right), \label{eq:series_Y} \\
X_\phi(\phi)   &= -\frac{1}{\tilde{c}} + \sum_{n=1}^{N-1} b_n \cos\left(\frac{kn}{\tilde{c}}\phi \right), \label{eq:series_XP}   \\
\tilde{Y}(\phi) &= \sum_{n=1}^{N-1} c_n \cos\left(\frac{kn}{\tilde{c}}\phi \right). \label{eq:series_Yt} 
\end{align}
We take $a_n$, $c_n$, $B_1$, $B_2$, $Q_1$, $Q_2$, $Q_3$, $h_0$, $g_0$, and $\tilde{c}$ as unknowns. This results in $2N+6$ unknowns. We differentiate (\ref{eq:series_Y}) and (\ref{eq:series_Yt}) with respect to $\phi$ to find $Y_\phi$ and $\tilde{Y}_\phi$. We then find $b_n$ by numerically satisfying the boundary integral equation (\ref{eq:int1}). The integrals are evaluated at midpoints, and approximated using the trapezoidal rule. The discretised system is then solved via Newton's method. Given $X_\phi$, $Y_\phi$, and $Y$, one can recover $h_\phi$ explicitly using equation (\ref{eq:h}). Expressing $h_\phi$ explicitly in other unknown variables (and hence reducing the number of unknowns in the nonlinear solver) was inspired by the work of \cite{wang2014numerical}, who deployed such a trick for two-layer hydroelastic interfacial waves with a free-surface. We integrate $h_\phi$ numerically to find $h$, and then compute $\tilde{X_\phi}$ at midpoints explicitly by re-arranging equation (\ref{eq:int3}). A four-point interpolation formula  is used to approximate values of $\tilde{X_\phi}$ at the meshpoints $\phi_I$. One can then compute $g_\phi$ explicitly using (\ref{eq:g}), and then integrate for values of $g$ at each meshpoint. It is left then to solve the integral equations (\ref{eq:int2}) and (\ref{eq:int4}). Satisfying these equations at midpoints results in $2N$ equations. Another two equations are obtained by satisfying the far-field equations \eqref{eq:hfarfield} and \eqref{eq:gfarfield}. These equations ensure that any ESW computed with this numerical scheme had a far-field given by a uniform stream with speeds $-c$ in each layer. Another three equations are obtained by fixing $H_i$ for each layer.
We do this by discretising $\psi_i$ in each layer, and then evaluating values of $x_{\phi}$ at the first midpoint $\phi_1^M$. We then compute $H_i$ using the identities
 \begin{equation}
 H_3 = \int\displaylimits_0^{Q_3} x_{\phi}(\phi_1^M,\psi_3) \, \mathrm{d} \psi_3, \hspace{0.3cm}  H_2 = \int\displaylimits_0^{Q_2} \frac{x_{\phi}(\phi_1^M,\psi_2)}{h_\phi(\phi_1^M)} \, \mathrm{d} \psi_2,  \hspace{0.3cm}  H_1 = \int\displaylimits_0^{Q_1} \frac{x_{\phi}(\phi_1^M,\psi_1)}{g_\phi(\phi_1^M)} \, \mathrm{d} \psi_1.
 \end{equation} 
Here, values of $x_\phi$ inside the flow domains (rather than on the boundaries) needs to be recovered using Cauchy's integral equation, resulting in equations similar to (\ref{eq:int1}).
 For an ESW, the far-field corresponds to a uniform stream with layers of depth $H_i$. A final equation is obtained by fixing something about the solution, such as some measure of the amplitude (for example, $Y(0)$), or the speed of the wave $\tilde{c}$ (\ref{eq:c1}). The system is solved via Newton's method, and we say a solution has converged when the $L_\infty$ norm of the residuals are of the order $O(10^{-10})$. 

As mentioned before, the above method results in a $2N+6$ system of unknowns and equations, which we solve via Newton's method. This is a much smaller system than the $8N+1$ unknowns found in the numerical scheme used by \cite{rusaas2002solitary}, and hence justifies our decision to solve the integrals in the potential space. 

We briefly mention here that when evaluating the integrals on the boundary, the singularity occurring in the integrals are handled by evaluating the integrals at midpoints. However, when seeking values inside the domains, such as in the evaluation of $H_i$, it is found that the singular nature of the integrals becomes problematic for points close to the boundary. To solve this, we rewrite the integrals with the singularity removed. For example, consider the integral of some analytic function $F$ on the annulus in the $t$-plane, given by (\ref{eq:t3}). Denoting the boundary of the annulus traversed counterclockwise as $C$, and let $t_0$ be some point inside the annulus which maps to ($\phi_0$,$\psi_0$) in $\Omega_3^f$. Then we write
\begin{equation}
F(t_0) = \frac{1}{2\pi i} \oint_C \frac{F(t)-F(t_1)}{t-t_0} + \frac{1}{2\pi i} \oint_C \frac{F(t_1)}{t-t_0},
\end{equation}
where $t_1=e^{-ki\phi_0/\tilde{c}}$. The first integral is evaluated numerically, where the singularity as $t_0\rightarrow t_1$ is removed by the new numerator. The second integral is evaluated analytically, which using Cauchy's integral formula gives simply $F(t_1)$. Hence, the computation is the same as in (\ref{eq:CIF}), except with a new correction term, given by the term in the square brackets below
\begin{equation}
F(t_0) = \frac{1}{2\pi i} \oint_C \frac{F(t)}{t-t_0} + \left[F(t_1) - \frac{F(t_1)}{2\pi i} \oint_C \frac{1}{t-t_0} \right].
\end{equation}
The correction term measures inaccuracies in evaluating the integral of a simple pole numerically when using the trapezoidal rule. The errors only become large near the boundary (that is, $t_0$ near $t_1$). It is found the above method improves the convergence of the evaluation of the integral as the number of meshpoints $N$ is increased.

\end{appendix}

\end{document}